\documentclass[floats,floatfix,showpacs,amssymb,prd,twocolumn,superscriptaddress,nofootinbib]{revtex4-1}

\usepackage{graphicx,epsf, epsfig, amssymb}
\usepackage{bm}
\usepackage{longtable}
\usepackage{color}
\usepackage[breaklinks]{hyperref}
\usepackage{amsfonts,amsmath,amssymb,mathrsfs}

\def\nn{\nonumber}
\def\be{\begin{equation}}
\def\ee{\end{equation}}
\def\beq{\begin{eqnarray}}
\def\eeq{\end{eqnarray}}

\newcommand{\om}{\omega}

\def\f{\frac}

\begin{document}

\title{Quasinormal ringing of Kerr black holes. \\II. Excitation by
  particles falling radially with arbitrary energy}

\author{Zhongyang Zhang\footnote{Electronic address:
    zzhang1@olemiss.edu}} \affiliation{Department of Physics and
  Astronomy, The University of Mississippi, University, MS 38677, USA}

\author{Emanuele Berti\footnote{Electronic address:
    berti@phy.olemiss.edu}} \affiliation{Department of Physics and
  Astronomy, The University of Mississippi, University, MS 38677, USA}

\affiliation{California Institute of Technology, Pasadena, CA 91109,
  USA}

\author{Vitor Cardoso\footnote{Electronic address:vitor.cardoso@ist.utl.pt}} 
\affiliation{CENTRA, Departamento de
  F\'{\i}sica, Instituto Superior T\'ecnico, Universidade T\'ecnica de
  Lisboa - UTL, Av.~Rovisco Pais 1, 1049 Lisboa, Portugal}
  \affiliation{Department of Physics and Astronomy, The University of
  Mississippi, University, MS 38677, USA}
  \affiliation{Perimeter Institute for Theoretical Physics Waterloo, Ontario N2J 2W9, Canada}

\begin{abstract}
The analytical understanding of quasinormal mode ringing requires an
accurate knowledge of the Green's function describing the response of
the black hole to external perturbations. We carry out a comprehensive
study of quasinormal mode excitation for Kerr black holes. Relying on
the formalism developed by Mano, Suzuki and Takasugi, we improve and
extend previous calculations of the quasinormal mode residues in the
complex frequency plane (``excitation factors $B_q$''). Using these
results we compute the ``excitation coefficients'' $C_q$ (essentially
the mode amplitudes) in the special case where the source of the
perturbations is a particle falling into the black hole along the
symmetry axis.  We compare this calculation with numerical
integrations of the perturbation equations, and we show quantitatively
how the addition of higher overtones improves the agreement with the
numerical waveforms. Our results should find applications in models of
the ringdown stage and in the construction of semianalytical template
banks for gravitational-wave detectors, especially for binaries with
large mass ratios and/or fast-spinning black holes.
\end{abstract}


\maketitle
\section{Introduction}
Distorted black holes (BHs) emit gravitational radiation. A spectral
decomposition of the perturbation response of the Schwarzschild
\cite{Leaver:1986gd} and Kerr \cite{Berti:2006wq} geometries using
Green's function techniques shows that a discrete sum of quasinormal
modes (QNMs) -- damped oscillations whose frequencies and damping
times depend only on the BH mass and angular momentum -- will
dominate the response at all but very early and very late times.
Because of the qualitative similarity with a ringing bell, this
intermediate stage is known as ``ringdown'' in the gravitational-wave
literature \cite{Kokkotas:1999bd,Nollert:1999ji,Berti:2009kk}.

Numerical simulations show that binary BH mergers in general
relativity inevitably result in the formation of a distorted rotating
remnant, which radiates ringdown waves while settling down into a
stationary (Kerr) solution of the Einstein equations in vacuum.
Despite the great advances in binary BH simulations in four
\cite{Pretorius:2007nq,Sperhake:2011xk,Pfeiffer:2012pc} and higher
dimensions \cite{Sperhake:2013qa}, the excitation of the QNMs of the
remnant BH resulting from a merger is still poorly
understood. Perturbative techniques are especially valuable to
understand ringdown excitation in situations that pose a particular
challenge to numerical simulations, namely:

\noindent
1) {\em Large mass-ratio binaries.} One of the frontiers in numerical
simulations of BH mergers are quasicircular binaries with large mass
ratios. Progress in this direction has been slow but steady, both in
the quasicircular case -- where initial record mass ratios
$q=m_1/m_2=10$ \cite{Gonzalez:2008bi} have been broken using
``hybrid'' techniques \cite{Lousto:2010qx,Nakano:2011pb} -- and in the
head-on case, where simulations with $q=100$ have recently been
performed using different approaches
\cite{Sperhake:2011ik,East:2013iwa}. In this regime, perturbation
theory is crucial to validate and/or optimize numerical simulations.

\noindent
2) {\em Large spins.} Numerical simulations of BH binaries are usually
carried out using either the Baumgarte-Shapiro-Sasaki-Nakamura (BSSN)
formulation of the Einstein equations and a finite-difference scheme,
or using the harmonic formulation and spectral methods. The first
class of simulations is limited to dimensionless spins
$a/M=J/M^2\lesssim 0.93$, because this is the maximum spin that can be
achieved with puncture initial data \cite{Lousto:2012es}. Initial data
with spins as large as $a/M\sim 0.97$ can be constructed
\cite{Lovelace:2008tw} and have been evolved using spectral codes
\cite{Lovelace:2010ne,Lovelace:2011nu}. These simulations present a
significant challenge for modeling efforts using effective-one-body
techniques when one considers binaries with aligned spins $a/M\gtrsim
0.7$ \cite{Taracchini:2012ig}. Models of the late merger and ringdown
phase can be significantly improved by using first-principle
calculations in BH perturbation theory, rather than a phenomenological
matching of inspiral waveforms with QNM superpositions of largely
arbitrary amplitudes and starting times.

\noindent
3) {\em Higher dimensions.} Numerical simulations in higher dimensions
are very challenging, and simple calculations in BH perturbation
theory can give insight into the results of the simulations. For
example, the qualitative behavior of the energy and linear momentum
radiated by particles falling into higher-dimensional
Schwarzschild-Tangherlini BHs (predicted in
\cite{Berti:2003si,Berti:2010gx}) is in excellent agreement with the
first numerical simulations in $D=5$ \cite{Witek:2010az}: see
e.g. \cite{Yoshino:2011zz,Cardoso:2012qm} for reviews.

First-principle calculations of QNM excitation in four space-time
dimensions would be particularly beneficial in building semianalytical
models of the merger/ringdown phase, to be used as matched-filtering
templates in gravitational-wave searches. Here we carry out these
calculations in four spacetime dimensions considering, for simplicity,
head-on particle infalls into Schwarzschild and Kerr BHs. Our study
improves and extends the results of \cite{Berti:2006wq}.

\subsection{Excitation factors and excitation coefficients \label{sec:excitation}}
The gravitational radiation from a perturbed Kerr BH is usually
described in terms of the Weyl scalar $\psi_4$
\cite{Teukolsky:1972my,Teukolsky:1973ap}, which can be decomposed in
different multipolar components (say $\psi_{lm}$) by using
spin-weighted spheroidal harmonics with angular indices $(l\,,m)$ (see
e.g.~\cite{Berti:2005gp}). In the ringdown stage, each $\psi_{lm}$ can
be expressed as a sum of complex exponentials: schematically,
\beq 
\psi_{lm}\sim 
\sum_{n=0}^{\infty}C_{lmn} \exp
\left[-i\om_{lmn}(t-r_*)\right]
\,,
\eeq
where the frequencies $\om_{lmn}$ are complex, $t$ denotes time as
measured by an observer at infinity, $r_*$ is a radial ``tortoise''
coordinate, and the index $n$ (``overtone index'') sorts the modes by
increasing imaginary part ($n=0$ corresponding to the smallest
imaginary part and to the longest damping time).  To simplify the
notation, we will sometimes replace the indices $(l\,,m\,,n)$ by a
collective index $q$. 

The problem of extracting the QNM contribution to a generic signal was
first studied in detail by Leaver~\cite{Leaver:1986gd}.
The complex amplitudes $C_q$ of each complex exponential, also called
``excitation coefficients'', depend on the source of the perturbation
(see e.g.~\cite{Nollert:1998ys,Berti:2006hb,Dorband:2006gg}).  The
excitation coefficients can be factorized into the product $C_q=B_q
I_q$ of a source-independent ``excitation factor'' $B_q$ and of a
source-dependent integral $I_q$.  The integral $I_q$ is in general
divergent, but it can be regularized, yielding a finite answer in
agreement with other perturbative calculations
\cite{Leaver:1986gd,Sun:1988tz,Hadar:2009ip,Hadar:2011vj}.

To illustrate the origin of this factorization, consider the following
prototypical ODE governing arbitrary perturbations around a BH.  The
perturbation is characterized by a wave function $\Psi$ with source
$Q$ (representing for example the perturbation due to infalling
matter):
\beq  
\frac{\partial^2}{\partial r_*^2}\Psi-\frac{\partial^2}{\partial t^2}\Psi-V\Psi=-Q(t)\,,
\eeq
where $r_*$ is a radial ``tortoise coordinate'', and the potential
$V=V(r_*)$.  The wave function $\Psi$ can describe curvature-related
quantities in the formalism by Sasaki and Nakamura
\cite{Sasaki:1981sx} and it is directly related to metric
perturbations in the Regge-Wheeler/Zerilli formalism
\cite{Regge:1957rw,Zerilli:1971wd}.

The QNM contribution to the time-domain Green's function $G_Q$ reads
\beq
\Psi_Q(r_*,t)=
\int^\infty_{-\infty}\int^\infty_{-\infty}G_Q(r_*,t|r_*',t')
Q(r_*',t')dr'_*dt'\,,
\nn
\eeq
%
where (see e.g. \cite{Leaver:1986gd})
\beq\label{GQ}
&&G_Q(r_*,t|r_*',t')=\nn\\
&=&2\text{Re}\left[\sum_{q=0}^{\infty}
B_q \psi_q(r_*)\psi_q(r_*')e^{-i\om_q(t-t'-r_*-r'_*)}
\right]\,.
\eeq

The coefficients $B_q$ are the (source-independent) {\it excitation
  factors}, and $\psi_q(r)$ denotes solutions of the homogeneous
equation normalized such that $\psi_q(r)\to 1$ as $r_*\to \infty$ .
It is convenient to introduce also the source-dependent {\it
  excitation coefficients} $C_q$:
\be
\label{Cq}
C_q=B_q I_q\,,\\
\ee
where
\be
I_q\equiv \int_{-\infty}^\infty
e^{i\om_qr_*'}\psi_q(r_*')q(r_*',\om)dr_*'\,,\label{eq:source_def}
\ee
and where the frequency-domain source term is
\beq
q(r_*',\om)=\int^\infty_{-\infty} e^{i\om t'}Q(r_*',t')dt'\,. 
\eeq
The calculation of the $C_q$'s involves an integral in $r_*$
from the horizon ($r_*=-\infty$) out to spatial infinity
($r_*=\infty$). The integral usually diverges at the horizon; one of
the proposed methods to eliminate this divergence is discussed below in Section \ref{sec:regularization}.
With these definitions, the ringdown waveform can be written as:
\beq
\Psi(r_*,t)=2\text{Re}\left[\sum_{q=0}^{\infty}C_q \psi_q(r_*)
e^{-i\om_q(t-r_*)}\right]\,.
\eeq
As $r_*\to\infty$ we have $\psi_q(r_*)\to 1$, so that
\beq
\label{semianal}
\Psi_Q(r_*\to\infty,t)=2\text{Re}\left[\sum_{q=0}^{\infty}C_q
 e^{-i\om_q(t-r_*)}\right]\,.
\eeq

To summarize, the complex excitation factors $B_q$ are a ``universal''
intrinsic property of the BH which describes the excitability of each
mode, independently of the source of the excitation. On the other
hand, the complex excitation coefficients $C_q$ are related to the
amplitude of each QNM in response to a {\em specific} source inducing
the oscillations.

\subsection{Plan of the paper}
%
In the first part of this paper (Section \ref{sec:Bq}) we compute a
catalog of QNM excitation factors $B_q$ for Kerr BHs using the
formalism developed by Mano, Suzuki and Takasugi
(\cite{Mano:1996vt,Mano:1996mf}, henceforth MST). By using this
technique we confirm and extend results obtained some years ago by two
of us \cite{Berti:2006wq}. The main advantage of the MST method is
that it does not require the (generally nontrivial) evaluation of
Coulomb wave functions, which was instead necessary in
\cite{Berti:2006wq}. This allows us to produce accurate tables of the
$B_q$'s for the modes that are most interesting in gravitational-wave
detection (multipolar indices $l\leq 7$ and overtone indices
$n=0,\,\dots,\,4$). These tables (and similar tables for perturbations
of spin $s=0$ and $s=1$) will be made publicly available on a website,
along with a {\sc Mathematica} notebook that can be adapted to
generate further tables if necessary \cite{rdweb}.

In the second part of the paper we compute the excitation coefficients
$C_q$ for a classic problem in perturbation theory: the calculation of
the gravitational radiation emitted by particles falling into the
BH. We generalize work carried out by Leaver more than 25 years ago
\cite{Leaver:1986gd} (see also \cite{Sun:1988tz}). Whereas Leaver
considered only infalls from rest into a Schwarzschild BH, we present
detailed comparisons between numerical waveforms and excitation
coefficient calculations for particles falling with arbitrary energy
into Schwarzschild BHs (Section \ref{sec:Schw}) and we also consider
the case where the BH is rotating (Section \ref{sec:Kerr}). In
Section \ref{sec:Conclusions} we summarize our findings and point out
possible directions for future work. Appendix \ref{app:reg} gives
details about the regularization of divergent integrals in both the
Schwarzschild and Kerr cases.
In the whole paper we use geometrical units ($G=c=1$).

\section{Excitation factors in the Mano-Suzuki-Takasugi formalism}
\label{sec:Bq}

In this Section we present a detailed calculation of the excitation
factors $B_q$ for Kerr QNMs. We follow the MST formalism
(\cite{Mano:1996vt}; see also \cite{Mano:1996mf,Mano:1996gn}) and we
refer to the original papers for a more organic presentation of the
material; our intention here is to give a practical guide to the
calculation of the $B_q$'s within this formalism. The method is
different from -- but equivalent to -- Leaver's method
\cite{Leaver:1986gd}, that was used by two of us in
\cite{Berti:2006wq}. The main advantage of the MST formalism over
Leaver's method is that the MST formalism does not require any
(cumbersome) {\it evaluation} of Coulomb wave functions, as in
Leaver's original treatment, but only a {\em matching} of the
Coulomb-series expansion near infinity to an expansion in terms of
hypergeometric functions near the horizon, which is simpler to perform
in practice.

We will compute the excitation factors in both the Teukolsky and
Sasaki-Nakamura formalisms (see \cite{Berti:2006wq} for a
discussion). To begin with, let us define some quantities that will be
used below:
\beq r_\pm&=&M\pm\sqrt{M^2-a^2}\,, \quad \kappa=\sqrt{1-j^2}\,,
\nn\\
x&=&\frac{\omega(r_+-r)}{\epsilon\kappa}\,, \tau=\frac{\epsilon-a
m/M}{\kappa}\,, \quad \epsilon_\pm=\frac{\epsilon\pm\tau}{2}\,. 
\eeq
From now on we follow Leaver's conventions and set $2M=1$ (where $M$
is the BH mass). In these units, the parameter $a\in[0,\,1/2]$. In
order to make contact with the more usual $M=1$ units, we find it
convenient to introduce a second dimensionless spin parameter $j\equiv
2a\in[0,\,1]$.
For reference, intermediate results of our calculations for a specific
value of the spin ($a=0.4$, or $j=0.8$) are given in Table
\ref{tab:results}. In the remainder of this Section we will define and
compute the quantities listed in this Table.

\begin{center}
\begin{table*}[thb]
\begin{tabular}{cccc}
  \hline
  \hline
   & $s=-2$, $l=m=2$ & $s=-1$, $l=m=1$ & $s=0$, $l=m=2$ \\
  \hline
  $\omega_q$ & $1.172034 - 0.151259$i & $0.701679 - 0.152621$i & $1.41365 - 0.163041$i \\
  $A_{lm}$& $2.585294 + 0.205297$i & $1.67659 + 0.0810074$i & $5.95475 + 0.0106275$i \\
  $\nu$ & $-1.743843 - 0.701583$i & $-1.69028 - 0.320182$i & $-1.8012 - 0.0481726$i \\
  $a_4^\nu$ & $-1.32616\times10^{-3} + 1.43416\times10^{-3}$i & $-4.04792\times10^{-3} + 3.01211\times10^{-3}$i & $-0.229461 - 0.0295086$i \\
  $a_{-4}^{\nu}$ & $-4.52814\times10^{-3} - 2.12986\times10^{-2}$i & $8.02490\times10^{-5} -2.25538\times10^{-4}$i & $1.47272\times10^{-3} +3.40832\times10^{-4}$i \\
  $K_\nu$ & $1.06144\times10^{-3} + 7.43631\times10^{-4}$i & $-0.0812872 + 0.0682523$i & $-12.0419 + 1.20138$i \\
  $K_{-\nu-1}$ & $-8.19837\times10^{-2} - 9.20267\times10^{-1} $i & $1.55992 + 1.23780$i & $18.6581 + 3.85088$i \\
    $B_{lm\omega}^{\text{inc}}$ & $-2.80111\times10^{-16} + 3.11473\times10^{-16}$i & $-4.51443\times10^{-15} - 2.05141\times10^{-15}$i & $6.08313\times10^{-14} + 1.91604\times10^{-14}$i \\
    $B_{lm\omega}^{\text{ref}}$ & $3.16122 + 1.25413$i & $1.59262 - 0.363221$i & $1.27738 + 0.760771$i \\
    $B_{lm\omega}^{\text{trans}}$ & $15.4151 + 11.0126$i & $3.32227 + 0.409647$i & $0.496587 + 1.24305$i \\
      $\alpha_q^{\rm T}$ & $0.114759 - 0.241821$i & $-1.25046 - 1.01565$i & $-0.154117 - 3.58899$i \\
      $B_q^{\rm T}$ & $-0.240807 + 0.150102$i & $-0.153477 - 0.144681$i & $-0.0955564 + 0.0516867$i \\
      $B_q^{\rm SN}$ & $-0.0911231 + 0.0613455$i & $-0.0298959 - 0.119248$i & $-0.0955564 + 0.0516867$i \\
  \hline
\end{tabular}
\caption{\label{tab:results}Some intermediate quantities necessary to
  compute the excitation factors for the fundamental mode ($n=0$) of a
  Kerr BH with $a=0.4$ (or $j=0.8$). The three columns refer to
  gravitational ($s=-2$) perturbations with $l=m=2$, electromagnetic
  ($s=-1$) perturbations with $l=m=1$, and scalar ($s=0$)
  perturbations with $l=m=2$.}
\end{table*}
\end{center}

\subsection{\label{wAlm}Computing \texorpdfstring{$\omega_q$}{wq} and \texorpdfstring{$A_{lm}$}{Alm}}
In the Teukolsky formalism, the perturbations of a Kerr BH
are described by the Newman-Penrose scalar
$\psi_4$,
%
%
which is related to solutions $\phi$ of the Teukolsky equation by
$\phi\equiv\rho^{-4}\psi_4$, where $\rho=(r-ia\cos\theta)^{-1}$.
By expanding in Fourier components
\beq
\rho^{-4}\psi_4=\frac{1}{2\pi}\sum_{l=|s|}^\infty\sum_{m=-l}^{l}\int
e^{-i\omega t+im\varphi}S_{lm\omega}(\theta)R_{lm\omega}(r)
d\omega\nn \eeq
and performing a separation of variables, one finds that the radial
function $R_{lm\omega}$ and the angular function $S_{lm}$ must satisfy
the following equations:

\begin{widetext}
\beq \label{r1} \Delta\frac{d^2
R_{lm\omega}}{dr^2}+(s+1)(2r-1)\frac{d
R_{lm\omega}}{dr}+V(r)R_{lm\omega}&=&T_{lm\omega}\,, \\
\label{theta1} \frac{d}{du}\left((1-u^2)\frac{d
S_{lm}}{du}\right)+\left[a^2\omega^2u^2-2a\omega s
u+s+A_{lm}-\frac{(m+su)^2}{1-u^2}\right]S_{lm}&=&0\,,
\eeq
\end{widetext}
where $u=\cos \theta$ and $T_{lm\omega}$ is the Fourier transform of
the stress-energy tensor after separation of the angular dependence.
The potential $V(r)$ is given by
\beq 
V(r)&=&\left\{(r^2+a^2)^2\omega^2-2am\omega
r+a^2m^2\right.\nn\\
&+&\left. is[am(2r-1)-\omega(r^2-a^2)]\right\}\Delta^{-1}\nn\\
&+&2is\omega r-a^2\omega^2-A_{lm}\,, 
\eeq
where $A_{lm}$ is the angular separation constant corresponding to the
angular eigenfunctions $S_{lm}$
(known as ``spin-weighted spheroidal harmonics''). 
The eigenfrequency $\omega_{lmn}=\omega_q$ and the angular eigenvalue
$A_{lm}$ are determined by imposing QNM boundary conditions on the
radial equation~(\ref{r1}) and regularity conditions on the angular
equation (\ref{theta1}): see e.g.~\cite{Berti:2009kk}.
The radial and angular equations are solved via a series solution
whose coefficients $b^r_n$ and $b^\theta_n$ satisfy three-term
recursion relations of the form

\beq \label{three term theta}\alpha_0^\theta
b^{(r,\theta)}_1+\beta_0^{(r,\theta)}
b^{(r,\theta)}_0&=&0\,,\notag\\
\alpha_n^{(r,\theta)} b^{(r,\theta)}_{n+1}+\beta_n^{(r,\theta)}
b^{(r,\theta)}_n+\gamma_n^{(r,\theta)} b^{(r,\theta)}_{n-1}&=&0 \,, \eeq
where the superscript ($r$ or $\theta$) denotes association with the
radial or angular equation, and the coefficients of the three-term
recursion relations can be found in \cite{Leaver:1985ax}.

By the principle of minimal solutions, the convergence of the series
obtained via the three-term recursion relations is guaranteed by two
continued fraction relations (one coming from the radial series
expansion, the other from the angular series expansion) of the form
\beq\label{theta-continue}
\beta_0^\theta=\frac{\alpha^\theta_{0}\gamma^\theta_1}{\beta^\theta_{1}-\frac{\alpha^\theta_{1}\gamma^\theta_{2}}{\beta^\theta_{2}-...}}\,,\\
\label{r-continue}
\beta_0^r=\frac{\alpha^r_{0}\gamma^r_1}{\beta^r_{1}-\frac{\alpha^r_{1}\gamma^r_{2}}{\beta^r_{2}-...}}\,.
\eeq
or by any of their inversions, which are analytically -- but not
numerically -- equivalent \cite{Leaver:1985ax}.

We now have two complex equations, (\ref{theta-continue}) and
(\ref{r-continue}), in two complex unknowns, $\omega_q$ and
$A_{lm}$. By solving these equations numerically we find the
eigenvalues listed in the first two rows of Table
\ref{tab:results}. Numerical practice shows that the $q$th inversion
index for the radial equation is best suited for numerical searches of
the $q$th overtone $\omega_q$.  Numerical experimentation (and
analytical arguments \cite{Berti:2005gp}) show that the optimal
inversion number to find the angular eigenvalue with the correct limit
as $a\to 0$, i.e.
\be
A_{lm}\to l(l+1)-s(s+1)\,,
\ee
is equal to $l-\max(|m|,|s|)$.

\subsection{Angular momentum parameter \texorpdfstring{$\nu$}{Function nu} and matching function \texorpdfstring{$K_{\nu}$}{Function Knu}}
The basic idea of the MST method is to (1) find a first independent
solution of the radial equation $R_0^\nu$ in terms of a series of
hypergeometric functions (which does not converge at spatial infinity)
with expansion coefficients proportional to $a_n^\nu$, cf. Eq.~(2.21)
of \cite{Mano:1996vt}; (2) consider Leaver's construction of a series
of Coulomb wave functions $R_C^\nu$ that is valid near infinity; (3)
notice that the two solutions are identical modulo a $\nu$-dependent
constant, i.e.
\be 
R_0^\nu=K_\nu R_C^\nu\,. 
\ee
The expansion coefficients $a_n^\nu$ and the matching condition depend
on an ``angular momentum'' parameter $\nu$ which appears in the
three-term recurrence relation
\beq\label{nu-recursion} 
\alpha_n^\nu a_{n+1}^\nu+\beta_n^\nu
a_n^\nu+\gamma_n^\nu a_{n-1}^\nu=0\,, 
\eeq
where
\beq \label{alpha}
\alpha_n^\nu&=&\frac{i\epsilon\kappa(n+\nu+1+s+i\epsilon)(n+\nu+1+s-i\epsilon)}{(n+\nu+1)(2n+2\nu+3)(n+\nu+1+i\tau)^{-1}}\,,\nn\\
\label{beta}
\beta_n^\nu&=&-\lambda-s(s+1)+(n+\nu)(n+\nu+1)+\epsilon^2\,,\notag\\
&+&\epsilon(\epsilon-mq)
+\frac{\epsilon(\epsilon-mq)(s^2+\epsilon^2)}{(n+\nu)(n+\nu+1)}\,,\nn\\
\label{gamma}
\gamma_n^\nu&=&-\frac{i\epsilon\kappa(n+\nu-s+i\epsilon)(n+\nu-s-i\epsilon)}{(n+\nu)(2n+2\nu-1)(n+\nu-i\tau)^{-1}}\,.
\eeq\\
and $\lambda$ is related to the separation constant $A_{lm}$ by
\be
\label{eq:lambda}
\lambda=A_{lm}+(a\omega)^2-2am\omega\,.
\ee

The solution of the above recursion relation is ``minimal'' (i.e.,
the $a_n^\nu$'s give rise to a convergent series) if
\beq \label{nu}
\beta^\nu_0=\frac{\alpha^\nu_{-1}\gamma^\nu_0}{\beta^\nu_{-1}-\frac{\alpha^\nu_{-2}\gamma^\nu_{-1}}{\beta^\nu_{-2}-...}}
+\frac{\alpha^\nu_{0}\gamma^\nu_1}{\beta^\nu_{1}-\frac{\alpha^\nu_{1}\gamma^\nu_{2}}{\beta^\nu_{2}-...}}\,.
\eeq
This condition is only satisfied by a discrete set of (complex)
values of $\nu$.
Different inversions of Eq.~(\ref{nu}) yield different values of
$\nu$: for example, we could consider the first inversion
\beq\label{1}
\beta^\nu_1=\frac{\alpha^\nu_{0}\gamma^\nu_1}{\beta^\nu_{0}-\frac{\alpha^\nu_{-1}\gamma^\nu_{0}}{\beta^\nu_{-1}-\frac{\alpha^\nu_{-2}\gamma^\nu_{-1}}{\beta^\nu_{-2}-...}}
}+\frac{\alpha^\nu_{1}\gamma^\nu_2}{\beta^\nu_{2}-...}\eeq
or even a sequence of ``negative'' inversions, such as
\beq\label{-1}
\beta^\nu_{-1}=\frac{\alpha^\nu_{-2}\gamma^\nu_{-1}}{\beta^\nu_{-2}-...}
+\frac{\alpha^\nu_{-1}\gamma^\nu_0}{\beta^\nu_{0}-\frac{\alpha^\nu_{0}\gamma^\nu_{1}}{\beta^\nu_{1}-\frac{\alpha^\nu_{1}\gamma^\nu_{2}}{\beta^\nu_{2}-...}}}\,.
\eeq

Inversions are useful also for the radial and angular continued
fractions, but the numerical calculation of $\nu$ is a little
trickier: the numerical root $\nu$ can be different for different
inversions of the continued fraction, but this does not affect the
physics of the problem. The reason is that the eigenvalues $\nu$ have
the following properties:
(i) $\nu$ has period equal to 1: if $\nu$ is a solution, $\nu\pm1$
is also a solution;
(ii) If $\nu$ is a solution, $-\nu$ is also a solution.

Given the eigenvalue $\nu$ (as listed, e.g., in the third row of Table
\ref{tab:results}), it is straightforward to build up the series
coefficients $a_n^\nu$ from the three-term recursion relation
(\ref{nu-recursion}). If we choose the arbitrary normalization
constant such that $a_0^\nu=1$, we get (for example) the values of $a_4^\nu$
and $a_{-4}^\nu$ listed in rows four and five of Table \ref{tab:results}.

An important property of these coefficients is that
$a_{-n}^{-\nu-1}=a_n^\nu$: this can be shown starting from the
three-term recursion relation (\ref{nu-recursion}), and using
Eqs.~(\ref{gamma}).
Therefore we can denote them by $a_n^\nu$ when they refer to $K_\nu$,
and by $a_{-n}^{-\nu-1}$ when they refer to $K_{-\nu-1}$.

As we will see below, to obtain the QNM excitation coefficients we
must compute $K_\nu$ and $K_{-\nu-1}$, given by Eq.~(165) in
\cite{Sasaki:2003xr}:
\begin{widetext}
\beq K_\nu&=&\frac{e^{i\epsilon
\kappa}(2\epsilon\kappa)^{s-\nu-p}2^{-s}i^p\Gamma(1-s-2i\epsilon_+)\Gamma(p+2\nu+2)}
{\Gamma(p+\nu+1-s+i\epsilon)\Gamma(p+\nu+1+i\tau)\Gamma(p+\nu+1+s+i\epsilon)}
\times\left(\sum_{n=-\infty}^p\frac{(-1)^n}{(p-n)!(p+2\nu+2)_n\frac{(\nu+1+s-i\epsilon)_n}{(\nu+1-s+i\epsilon)_n}}a_n^\nu\right)^{-1}
\notag
\\
&\times&\left(\sum_{n=p}^{\infty}\frac{\Gamma(n+p+2\nu+1)}{(-1)^n(n-p)!}
\frac{\Gamma(n+\nu+1+s+i\epsilon)}{\Gamma(n+\nu+1-s-i\epsilon)}
\frac{\Gamma(n+\nu+1+i\tau)}{\Gamma(n+\nu+1-i\tau)}a_n^\nu\right)\,,
\eeq
\end{widetext}
where the notation $(x)_n$ is a shorthand for the following function
of $x$:
\be\label{sigurros} (x)_n\equiv \frac{\Gamma(x+n)}{\Gamma(x)}\,,
\ee
and $p$ can be any integer.
Both $K_\nu$ and $K_{-\nu-1}$ are independent of the choice of $p$;
indeed, this property can be used as a check of the
calculation. Representative values of $K_\nu$ and $K_{-\nu-1}$ are
listed in Table \ref{tab:results}.

\subsection{Amplitudes \texorpdfstring{$B_{lm\omega}^{\text{inc}}$}{Binc},
\texorpdfstring{$B_{lm\omega}^{\text{ref}}$}{Bref} and
\texorpdfstring{$B_{lm\omega}^{\text{trans}}$}{Btrans} in the Teukolsky formalism}
According to Eqs.~(167), (168) and (169) in \cite{Sasaki:2003xr}, the
ingoing-wave radial solution has the asymptotic behavior

\beq \label{RinAsy} R^{\text{in}}_{lm\omega}\to \left\{
\begin{array}{l}
B_{lm\omega}^{\text{trans}}\Delta^2e^{-ikr^*}~{\rm as}~r\to r_+\,,\\
r^3B_{lm\omega}^{\text{ref}}e^{i\omega
r^*}+r^{-1}B_{lm\omega}^{\text{inc}}e^{-i\omega r^*}~{\rm as}~r\to +\infty\,,\\
\end{array}
\right. \eeq
where the amplitudes are defined as:
\beq
B^{\text{inc}}_{lm\omega}&=&\omega^{-1}\left(K_\nu-ie^{-i\pi\nu}\frac{\sin\pi(\nu-s+i\epsilon)}{\sin\pi(\nu+s-i\epsilon)}K_{-\nu-1}\right)A_+^\nu\notag\\
&\times&e^{-i~\epsilon \ln \epsilon}\,,\\
B^{\text{ref}}_{lm\omega}&=&\omega^{-1-2s}\left(K_\nu+ie^{i\pi\nu}K_{-\nu-1}\right)A_-^\nu
e^{i~\epsilon \ln \epsilon}\,,\\
B^{\text{trans}}_{lm\omega}&=&\left(\frac{\epsilon\kappa}{\omega}\right)^{2s}
e^{i\epsilon_+\ln \kappa} \sum_{n=-\infty}^{\infty}a_n^\nu\,,
\eeq
and
\beq
A_+^\nu&=&e^{-(\pi/2)\epsilon}e^{(\pi/2)i(\nu+1-s)}2^{-1+s-i\epsilon}\notag\\
&\times&\frac{\Gamma(\nu+1-s+i\epsilon)}{\Gamma(\nu+1+s-i\epsilon)}\sum_{n=-\infty}^\infty
a_n^\nu\,,
\\
A_-^\nu&=&e^{-(\pi/2)\epsilon}e^{-(\pi/2)i(\nu+1+s)}2^{-1-s+i\epsilon}\notag\\
&\times&\sum_{n=-\infty}^\infty (-1)^n
\frac{(\nu+1+s-i\epsilon)_n}{(\nu+1-s+i\epsilon)_n} a_n^\nu\,. \eeq

The QNM boundary conditions require that
$B_{lm\omega}^{\text{inc}}$ must vanish at the QNM frequencies
$\omega_q$. Table \ref{tab:results} shows that this indeed happens
within an accuracy very close to machine precision. The table also
lists reference values for $B_{lm\omega}^{\text{ref}}$ and
$B_{lm\omega}^{\text{trans}}$.

\subsection{\texorpdfstring{$\alpha_q^{\rm T}$}{alphaqT} in the Teukolsky formalism}

The excitation factors (in the Teukolsky formalism) are defined as
\beq\label{BqT}
B^{\rm T}_q=-\frac{A^{\rm T}_{\text{out}}(\omega_q)}{2i\omega_q\alpha_q^{\rm T}}\,.
\eeq
Here
\be \alpha_q^{\rm T}\equiv
i\left(\frac{dA^{\rm T}_{in}}{d\omega}\right)_{\omega_q}\,, \ee
and furthermore
\beq A^{\rm T}_{\text{in}}\equiv
\frac{B_{lm\omega}^{\text{inc}}}{B_{lm\omega}^{\text{trans}}}\,,
\quad A^{\rm T}_{\text{out}}\equiv
\frac{B_{lm\omega}^{\text{ref}}}{B_{lm\omega}^{\text{trans}}}\,.
\eeq

Note that we can divide both $B_{lm\omega}^{\text{inc}}$ and
$B_{lm\omega}^{\text{ref}}$ by some arbitrary function $G(\omega)$
without affecting the excitation factors $B_q^{\rm T}$.  This is because
$B_{lm\omega}^{\text{inc}}$ must vanish at the QNM frequencies
$\omega_q$, so $G(\omega)$ is just an arbitrary rescaling (or
normalization) factor. The proof is trivial:
\beq
B_q^{\rm T}\propto\left(\frac{B_{lm\omega}^{\text{ref}}}{dB_{lm\omega}^{\text{inc}}/d\omega}\right)_{\omega_q}
=\left(\frac{B_{lm\omega}^{\text{ref}}/G}{d[B_{lm\omega}^{\text{inc}}/G]/d\omega}\right)_{\omega_q}
\eeq

The simplest choice would be to set $G=1$, but in order to reproduce
all of the values listed in Leaver's Table I \cite{Leaver:1986gd},
especially $\alpha^{\text{SN}}_q$ and $A^{\text{SN}}_{\text{out}}$, we
choose a normalization factor
\be 
G=B_{lm\omega}^{\text{trans}}\,. 
\ee

To get $\alpha_q^{\rm T}$ we must compute the derivative of
$A^{\rm T}_{\text{in}}$ with respect to $\omega$. We first compute
$A^{\rm T}_{\text{in}}$ at the QNM frequency $\omega_q$,
$A^{\rm T}_{\text{in}}(\omega_q)$. Then we consider a new frequency
$\omega_q+\delta$, and we repeat the calculation described above to
get $A^{\rm T}_{\text{in}}(\omega_q+\delta)$; note in particular that when
we repeat the first step (as described in Section \ref{wAlm}) we use
the angular continued fraction to obtain a ``new'' angular constant,
evaluated at $\omega_q+\delta$. Finally we can compute the
derivative by finite differencing:
\be \alpha_q^{\rm T}=
i\frac{A^{\rm T}_{\text{in}}(\omega_q+\delta)-A^{\rm T}_{\text{in}}(\omega_q)}{\delta}\,.
\ee

In our calculation we set $\delta=10^{-7}$ (i.e. we differentiate
along the real axis); as a check of our finite-differencing procedure
we also repeat the calculation with $\delta=10^{-7}i$ (i.e.,
differentiating along the pure-imaginary axis). The two results
usually agree to better than one part in $10^{6}$.

\subsection{Excitation factors in the Teukolsky (\texorpdfstring{$B^{\rm T}_q$}{Function BqT})
and Sasaki-Nakamura (\texorpdfstring{$B^{\text{SN}}_q$}{Function BqSN}) formalisms}
The excitation factors in the Teukolsky formalism were defined in
Eq.~(\ref{BqT}). It turns out that for many practical purposes,
including the calculation of radiation from infalling point particles
that will be presented later on in this paper, it is more convenient
to use the Sasaki-Nakamura wave function $X$, related to Teukolsky's
by
\be \label{XToR}
X=\frac{\sqrt{r^2+a^2}}{\Delta}\left (\alpha(r)R+\frac{\beta(r)}{\Delta}R' \right ) \,,
\ee
where the prime stands for a derivative with respect to
$r$. Specializing to the case presented in Appendix B of Sasaki and
Nakamura \cite{Sasaki:1981sx} [i.e, $f=h=1$ and $g=(r^2+a^2)/r^2$],
the functions $\alpha$ and $\beta$ are, respectively:
\beq \alpha&=&-\frac{iK}{\Delta^2}\beta+3iK'+\lambda+ \frac{6\Delta}{r^2}\,,\\
\beta&=&\Delta\left [-2iK+\Delta'-4\frac{\Delta}{r}\right ]\,. \eeq
Here $K=(r^2+a^2)\omega-am$, $\Delta=r^2-2Mr+a^2$ and $\lambda$ was
defined in Eq.~(\ref{eq:lambda}). Then the Sasaki-Nakamura wave
function $X$ satisfies
\be \frac{d^2X}{dr_*^2}-{\cal F}\frac{dX}{dr_*}-{\cal U}X={\cal S}\,,\label{eq:SN} \ee
where the tortoise coordinate is defined as
$\frac{dr}{dr_*}=\frac{r^2+a^2}{\Delta}$.  The tortoise coordinate is
defined up to an integration constant, which we fix once and for all
by setting
\be
r_*=r+\frac{2M r_+}{r_+-r_-}\log{(r-r_+)}-\frac{2Mr_-}{r_+-r_-}\log{(r-r_-)}\,.
\ee

The functions ${\cal F}$ and ${\cal U}$ are given by
\beq 
\nn
{\cal F}&=&\frac{\Delta}{r^2+a^2}F\,,\quad F\equiv \frac{\gamma'}{\gamma}\,,
\\
\nn
\gamma&\equiv&\alpha \left (\alpha+\frac{\beta'}{\Delta}\right )-\frac{\beta}{\Delta}
\left (\alpha'-\frac{\beta}{\Delta^2}V\right )\,,\label{defgamma}\\
{\cal U}&=&\frac{\Delta U}{(r^2+a^2)^2}+G^2+\frac{dG}{dr_*}-\frac{\Delta G F}{r^2+a^2}\,,\nn\\
\nn
G&\equiv&-\frac{\Delta'}{r^2+a^2}+\frac{r\Delta}{(r^2+a^2)^2}\,,\\ 
\nn
U&=&-V+\frac{\Delta^2}{\beta}\left [\left
((2\alpha+\frac{\beta'}{\Delta}\right)'-\frac{\gamma'}{\gamma}\left (\alpha+\frac{\beta'}{\Delta} \right )
\right ] \,.
\eeq
Note that our Teukolsky potential $V$ differs by an overall minus sign
from the potential used by Sasaki and Nakamura, and that
\be 
\lim_{r\to \infty}\gamma \equiv \gamma_{\infty}= 
\lambda (2+\lambda)-12iM\omega-12a\omega\left (\omega a-m\right )\,.
\ee
When $a\to 0$ the Sasaki-Nakamura potential reduces, by construction,
to the so-called Regge-Wheeler potential (cf.  Section~\ref{sec:Schw}
below for more details).  The asymptotic behavior of the
Sasaki-Nakamura wave function is
\beq X &\sim& A_{\rm trans} e^{-ikr_*}\,,\quad r\to r_+\,,\\
X&\sim& A_{\rm in} e^{-i\omega r_*}+ A_{\rm out} e^{i\omega r_*}\,,\quad r \to \infty\,.\eeq
where $k=\omega-am/r_+$, and the coefficients can be related to the
corresponding Teukolsky coefficients by
\beq 
A_{\rm in}^{\rm T}&=&-\frac{1}{4\omega^2}A_{\rm in}\,,\\ 
A_{\rm out}^{\rm T}&=&-\frac{4\omega^2}
{\lambda(\lambda+2)-6i\omega-12a\omega(a\omega-m)} A_{\rm out}\,, 
\label{relasympt}
\eeq
and $\lambda\equiv A_{lm}+(a\omega)^2-2am\omega$. The normalization
at the horizon is such that
\beq
A_{\rm trans}&=&r_+^{1/2} \left[ (8-12i\omega-4\omega^2)r_+^2\right.\\
&+&(12iam-8+8am\omega+6i\omega)r_+\nn\\ 
&-&\left. 4a^2m^2-6iam+2\right]\nn\,.
\eeq
A change of wave function of the form
\be X
=\exp\left[\int \f{{\cal F}}{2} dr_*\right] X_2=X_2\sqrt{\gamma}\,
\ee
eliminates the first derivative, yielding
\be \frac{d^2X_2}{dr_*^2}+ \left(\f{{\cal F}'}{2}-\f{{\cal F}^2}{4}-\cal U\right) X_2= {\cal S}\exp\left[-\int
\f{{\cal F}}{2} dr_*\right]=\frac{{\cal S}}{\sqrt{\gamma}}\,. \ee

To get the excitation factors in the Sasaki-Nakamura formalism we only
need the asymptotic relation between $X$ and $R$,
Eq.~\eqref{relasympt} (similar relations are presented in
\cite{Berti:2006wq} for scalar and electromagnetic
perturbations). Denoting scalar, electromagnetic and gravitational
perturbations by the subscript $0$, $-1$ and $-2$ respectively, and
dropping the ``$q$'' subscripts to simplify the notation, we have:
\beq
B_0^{\text{SN}}&=&B_0^{\rm T}\,,\\
B_{-1}^{\text{SN}}&=&
-\frac{2am\omega_q-A_{lm}-a^2\omega_q^2}{4\omega_q^2}B_{-1}^{\rm T}\,,\\
B_{-2}^{\text{SN}}&=&
\frac{\lambda(\lambda+2)-6i\omega_q-12a\omega_q(a\omega_q-m)}
{16\omega_q^4}B_{-2}^{\rm T}\,. 
\eeq

The results for $a=0.4$ ($j=0.8$) are listed in the last row of Table
\ref{tab:results}.  All of the $B_q$'s (for $s=0\,,-1\,,-2$) match the
results of Paper I, but now the computation does not involve tricky
evaluations of the Coulomb wave functions. This allows us to compute
excitation factors for a larger range of spin values, and for a larger
set of values of $(l\,,m)$ and of the overtone number $n$. An
extensive catalog of results for Kerr perturbations of spin $s=0$, $1$
and $2$, $l=s,\,\dots,7$ and $n=0,\,\dots,3$ is provided online in the
form of downloadable numerical tables \cite{rdweb}.

\section{Excitation factors and excitation coefficients for Schwarzschild black holes}
\label{sec:Schw}

\subsection{Excitation factors for the Zerilli and Regge-Wheeler equations}

Perturbations of rotating (Kerr) BHs are conveniently described using
curvature-related quantities in the Newman-Penrose approach. As
discussed in the previous section, this naturally leads to the
definition of the excitation factors in either the Teukolsky or
Sasaki-Nakamura formalism (the latter being more suitable to numerical
calculations, due to the short-range nature of the source term of the
Sasaki-Nakamura equation).

For the Schwarzschild BH geometry, a (perhaps more physically
transparent) direct metric perturbation treatment can be performed.
The perturbations separate in two sectors depending on their behavior
under parity: the axial (or odd) and polar (or even)
sector. Odd-parity metric perturbations can be found from the
Regge-Wheeler wave function $\Psi^{(-)}$, and even-parity
perturbations lead to the Zerilli equation for a single wave function
$\Psi^{(+)}$. In both cases the problem reduces to the solution of a
wave equation of the form
\beq  
\frac{\partial^2}{\partial r_*^2}\Psi^{(\pm)}-\frac{\partial^2}{\partial t^2}\Psi^{(\pm)}-V^{(\pm)}\Psi^{(\pm)}=-Q^{(\pm)}(t)\,.
\eeq
Defining $\lambda=(l-1)(l+2)/2$, the Zerilli potential reads
\beq
V^{(+)}=\left(\frac{r-1}{r}\right)
\frac{8\lambda^2(\lambda+1)r^3+12\lambda^2r^2+18\lambda
r+9}{r^3(2\lambda r+3)^2}\,,
\eeq
whereas the Regge-Wheeler potential reads
\beq
V^{(-)}=\frac{r-1}{r^3}\left[l(l+1)-\frac{3}{r}\right]\,.
\eeq

These equations can be solved in the frequency domain using the
approach followed by Leaver~\cite{Leaver:1986gd} and summarized
below. At the QNM frequencies, the Regge-Wheeler wave function,
normalized such that $\psi_q^{(-)}(r)\to 1$ as $r\to\infty$, reads:
\beq
\psi_q^{(-)}(r)&=&
\left(1-\frac{1}{r}\right)^{-2i\om_q}
\left[\sum_{n=0}^{\infty}a_n(\om_q)\right]^{-1}\nn\\
&\times&\left[\sum_{n=0}^{\infty}a_n(\om_q)(1-1/r)^n\right]\,,
\eeq
where the coefficients $a_n$ can be computed from a three-term
recursion relation (cf.~Appendix A in~\cite{Leaver:1986gd}).  A simple
relation between the homogeneous solutions of the Zerilli and
Regge-Wheeler equation was found by
Chandrasekhar~\cite{Chandrasekhar:1985kt} (see also Eqs.~(102)-(104)
in~\cite{Leaver:1986gd}). Using the Chandrasekhar transformation, we
find that the Zerilli wave function $\psi_q^{(+)}(r)$, again
normalized such that $\psi_q^{(+)}(r)\to 1$ as $r\to\infty$, is
\begin{widetext} \beq  \psi_q^{(+)}(r)=\frac{(1-1/r)^{-2i\om_q}}{\sum
a_n}\sum_{n=0}^\infty\left[ \left(1+\frac{-6i\om_q(2\lambda
r+3)+9(r-1)}{r^2(2\lambda
r+3)[2\lambda(\lambda+1)+3i\om_q]}+\frac{3n}{r^2[2\lambda(\lambda+1)+3i\om_q]}\right)
a_n \left(\frac{r-1}{r}\right)^n\right]\,.
\eeq
\end{widetext}
%

\begin{table*}[htb]
\begin{tabular}{ccccc}
  \hline
  \hline
  $B_q^{(-)}$ & $l=2$    &    $l=3$   & $l=4$   &  $l=5$  \\
  \hline
  $n=0$ & $0.126902 + 0.0203152$i     & $-0.0938898-0.0491928$i     & $0.065348 +0.0652391 $i     & $-0.0384465 - 0.0735239 $i \\
  $n=1$ & $0.0476826 -0.223755$i     & $-0.151135+0.269750$i     & $0.261488 -0.251524 $i     & $-0.363440 + 0.182660$i \\
  $n=2$ & $-0.190284+0.0157486$i     & $0.415029 +0.141038 $i      & $-0.549217-0.435328 $i     & $0.534171 + 0.828615 $i \\
  $n=3$ & $0.0808676 +0.0796126$i     & $-0.0434028-0.412747 $i     & $-0.316921+0.837911 $i     & $1.08630 - 1.14858 $i \\
  \hline
  \hline
  $B_q^{(+)}$ & $l=2$    &    $l=3$   & $l=4$   &  $l=5$  \\
  \hline
  $n=0$ & $0.120923 + 0.0706696$i     & $-0.0889796 - 0.0611757$i     & $0.0621266 + 0.069100$i     & $-0.0364029 - 0.0748073$i \\
  $n=1$ & $0.158645 - 0.253334$i     & $-0.191928 + 0.264820$i     & $0.279700 - 0.241825$i     & $-0.371542 + 0.173592$i \\
  $n=2$ & $-0.298933 - 0.0711341$i     & $0.436786 + 0.204560$i      & $-0.543211 - 0.478060$i     & $0.517754 + 0.854935$i \\
  $n=3$ & $0.113837 + 0.204137$i     & $-0.000920468 - 0.476365$i     & $-0.374502 + 0.859526$i     & $1.13916 - 1.14048$i \\
  \hline
\hline
\end{tabular}
\caption{\label{tab:Bq} Odd- and even-parity excitation factors for
  $l=2\,,3\,,4\,,5$.}
\end{table*}

As explained in Section~\ref{sec:excitation}, the QNM contribution to
the time-domain Green's function reads
\beq
\Psi_Q^{(\pm)}(r_*,t)=
\int^\infty_{-\infty}\int^\infty_{-\infty}G_Q^{(\pm)}(r_*,t|r_*',t')
Q^{(\pm)}(r_*',t')dr'_*dt'\,,
\nn
\eeq
with
\beq
&&G_Q(r_*,t|r_*',t')=\nn\\
&=&2\text{Re}\left[\sum_{q=0}^{\infty}
B_q^{(\pm)} \psi_q^{(\pm)}(r)\psi_q^{(\pm)}(r')e^{-i\om_q(t-t'-r_*-r'_*)}
\right]\,.\label{Green_Schw}
\eeq

Because the Sasaki-Nakamura wave function reduces to the Regge-Wheeler
wave function when $a\to 0$, the corresponding excitation factors are
related by
\be
B_q^{(-)}=B_{-2}^{\rm SN}(a=0)\,.
\ee
The even-parity excitation factors $B_q^{(+)}$ are related to the
odd-parity excitation factors $B_q^{(-)}$
by~\cite{Chandrasekhar:1985kt,Leaver:1986gd}
\beq
B_q^{(+)}=B_q^{(-)}\frac{2\lambda(\lambda+1)+3i\om_q}{2\lambda(\lambda+1)-3i\om_q}\,.
\eeq
Thus, one can compute excitation factors for both the Regge-Wheeler
and Zerilli representations using the excitation factors computed in
Section~\ref{sec:Bq}.

For completeness, in Table \ref{tab:Bq} we list the axial
($B_q^{(-)}$) and polar ($B_q^{(+)}$) Schwarzschild excitation factors
for the fundamental mode and for the first three overtones with
$l=2\,,3\,,4\,,5$. From Table \ref{tab:Bq} we see that the absolute
values of the excitation factors $|B_q^{(+)}|$ for different overtone
numbers $n$ and fixed $l$ are of comparable magnitude. Values of these
coefficients up to $l=7$ can be computed using the data available at
\cite{rdweb}.
%

%
\subsection{Excitation coefficients for low- and high-energy particle infalls}
\label{sec:excitation_coefficients}
We will now compute the source-dependent excitation {\it coefficients}
$C_q$ and compare them with actual waveforms for head-on infalls into
Schwarzschild or Kerr BHs along the symmetry axis.  This is a classic
problem addressed via the Regge-Wheeler-Zerilli formalism for
non-rotating BHs \cite{Davis:1971gg} and via the Sasaki-Nakamura
formalism for Kerr BHs \cite{Sasaki:1981sx}. The original analysis was
revisited by several authors, who considered particles falling with
generic energy and from finite distance into Schwarzschild BHs, Kerr
BHs, and higher-dimensional BHs
\cite{Ferrari:1981dh,Lousto:1996sx,Cardoso:2002ay,Cardoso:2002yj,Kodama:2003jz,Berti:2003si,Berti:2010ce,Berti:2010gx,Mitsou:2010jv}. In
four dimensions, head-on collisions with large mass ratio have even
become accessible to simulations in full numerical relativity
\cite{Sperhake:2011ik,East:2013iwa}.

In general, the source-dependent excitation coefficients $C_q^{(\pm)}$
are given by
\be
\label{Cq2}
C_q^{(\pm)}=B_q^{(\pm)} I_q^{(\pm)}\,,\\
\ee
where
\be\label{IqSchw}
I_q^{(\pm)}\equiv \int_1^\infty
e^{i\om_qr'}\psi_q^{(\pm)}(r')q^{(\pm)}(r',\om)(r'-1)^{i\om_q-1}r' dr'\,,
\ee
and where $q^{(\pm)}(r',\om)$ denotes the frequency-domain source term.
The calculation of the $C_q^{(\pm)}$'s involves an integral in $r$
from the horizon ($r=1$) out to spatial infinity ($r=\infty$). The
integral usually diverges at the horizon, but this divergence can be
eliminated, as discussed below.

For a four-dimensional Schwarzschild BH, radial infalls excite only
even (polar) perturbations and the source term in the Fourier domain
reads
%
\beq 
&&q(r,\om)=m_0 4\sqrt{2\pi}\sqrt{4l+2}\frac{r-1}{r(2\lambda
r+3)}\nn\\
&\times&\left[ 
\left(E^2-1+\frac{1}{r}\right)^{-1/2}+\frac{4E
\lambda}{i\om(2\lambda r+3)} \right]
e^{i\om T(r)}\,.
\label{source_generic}
\eeq
%
Here $m_0$ is the rest mass, $v_0$ is the speed of the particle at
spatial infinity, and $E=m_0/\sqrt{1-v_0^2}$ is the (conserved) energy
per unit mass of the infalling particle. For a particle falling from
rest at infinity, $E=1$; for a particle falling ultrarelativistically,
$E\to \infty$.

Since we work in perturbation theory, the amplitude of the radiation
is proportional to $m_0 E$, and therefore it is useful to define the
following rescaled quantities:
\beq 
\widetilde{C}_q=\frac{C_q^{(+)}}{m_0 E}\,,
\quad
\widetilde{I}_q=\frac{I_q^{(+)}}{m_0 E}\,. 
\eeq

The function $T(r)$ can be found by integrating the geodesic
equations, namely
\beq  
\frac{dT}{dr}=\frac{-r E}{(r-1)\sqrt{E^2-1+1/r}}\,. 
\eeq

\begin{table*}[thb]
\begin{tabular}{ccccc}
  \hline    
  \hline
  $E=1$ & $l=2$    &    $l=3$   & $l=4$   &  $l=5$  \\
  \hline
  $n=0$ & $-1.89425 - 0.906608$i     & $-0.184934 - 0.231572$i     & $-0.0178934 - 0.0566232$i     & $0.000637468 - 0.0141310$i \\
  $n=1$ & $-1.94463 - 0.521963$i     & $-0.226114 - 0.187532$i     & $-0.0288733 - 0.0511510$i     & $-0.00228156 - 0.0137320$i \\
  $n=2$ & $-2.02880 - 0.263614$i     & $-0.266489 - 0.148876$i      & $-0.0393956 - 0.0457755$i     & $-0.00509258 - 0.0132048$i \\
  $n=3$ & $-2.11182 - 0.115656$i     & $-0.306561 - 0.116049$i     & $-0.0496969 - 0.0405565$i     & $-0.00784699 - 0.0125698$i \\
  \hline
  \hline
  $E=10$ & $l=2$    &    $l=3$   & $l=4$   &  $l=5$  \\
  \hline
  $n=0$ & $-4.835573 + 0.874861$i     & $-1.195880 + 0.0709923$i     & $-0.449316 + 0.0101960$i     & $-0.209552 + 0.00308825$i \\
  $n=1$ & $-4.478522 + 0.683019$i     & $-1.177329 + 0.0667378$i     & $-0.446281 + 0.0112097$i     & $-0.208268 + 0.00284264$i \\
  $n=2$ & $-4.142391 + 0.502502$i     & $-1.156297 + 0.0606733$i      & $-0.443514 + 0.0110150$i     & $-0.207551 + 0.00277215$i \\
  $n=3$ & $-3.818084 + 0.354501$i     & $-1.134168 + 0.0540593$i     & $-0.440663 + 0.0104592$i     & $-0.206954 + 0.00269149$i \\
  \hline
\end{tabular}
\caption{\label{tab:Iq}Rescaled integrals $\widetilde{I}_q$ for
  $l=2\,,3\,,4\,,5$ for particle with energy $E=1$ (top) and $E=10$
  (bottom).}
\end{table*}

\begin{table*}[thb]
\begin{tabular}{ccccc}
  \hline
  \hline
  $E=1$ & $l=2$    &    $l=3$   & $l=4$   &  $l=5$  \\
  \hline
  $n=0$ & $-0.164989 - 0.243495$i     & $0.00228872 + 0.0319187$i     & $0.00280101 - 0.00475425$i     & $-0.00108031 + 0.000466722$i \\
  $n=1$ & $-0.440736 + 0.409836$i     & $0.0930598 - 0.0238868$i     & $-0.0204455 - 0.00732461$i     & $0.00323146 + 0.00470595$i \\
  $n=2$ & $0.587721 + 0.223120$i     & $-0.0859447 - 0.119540$i      & $-0.000483324 + 0.0436992$i     & $0.00865255 - 0.0111907$i \\
  $n=3$ & $-0.216793 - 0.444266$i     & $-0.0549996 + 0.146142$i     & $0.0534710 - 0.0275273$i     & $-0.0232746 - 0.00536977$i \\
  \hline
  \hline
  $E=10$ & $l=2$    &    $l=3$   & $l=4$   &  $l=5$  \\
  \hline
  $n=0$ & $-0.646559 - 0.235935$i     & $0.110752 + 0.0668420$i     & $-0.0286191 - 0.0304143$i     & $0.00785932 + 0.0155636$i \\
  $n=1$ & $-0.537460 + 1.242920$i     & $0.208289 - 0.324589$i     & $-0.122114 + 0.111058$i     & $0.0768867 - 0.0372097$i \\
  $n=2$ & $1.27404 + 0.144452 $i     & $-0.517466 - 0.210031$i      & $0.246188 + 0.206043$i     & $-0.109831 - 0.176008$i \\
  $n=3$ & $-0.507006 - 0.739057$i     & $0.0267961 + 0.540228$i     & $0.156039 - 0.382678$i     & $-0.232685 + 0.239092$i \\
  \hline
\end{tabular}
\caption{\label{tab:Cq}Rescaled excitation coefficients
  $\widetilde{C}_q$ for $l=2\,,3\,,4\,,5$ for particle with energy
  $E=1$ (top) and $E=10$ (bottom).}
\end{table*}
%

In order to compute the time-domain waveform generated by an infalling
particle, we first work in the frequency domain.  For a fixed (real)
frequency $\omega$, we integrate the homogeneous Zerilli equation
using a fourth-order accurate Runge-Kutta integrator. We use the
boundary condition that $\Psi^{(+)}\sim e^{-i\omega r_*}$ close to the
horizon and we integrate the homogeneous equation outwards up to some
large value of $r$. Starting from the numerically constructed
homogeneous solutions, we can use a Green's function technique to find
the solution of the inhomogeneous equation
\cite{Cardoso:2002yj,Cardoso:2002jr,Berti:2010ce}. Finally, we perform
an inverse Fourier transform to compute the time-domain wave function.

\subsection{Regularization at the horizon\label{sec:regularization}}
In order to find the excitation factors, one needs to evaluate
Eq.~(\ref{Cq2}) at the complex QNM frequency. At the horizon ($r\to
1$) the integrand appearing in the quantity $I_q^{(+)}$, as defined in
Eq.~(\ref{IqSchw}), can be written as a Frobenius series of the form
\beq 
e^{i\om_qr_*}\psi^{(+)}_q(r)q^{(+)}(r,\om_q) 
= \sum_{n=0}^\infty \xi_n (r-1)^{\zeta_q+n}\,.
\eeq
The convergent or divergent nature of the integral depends on the
value of $\zeta_q$, which in turn is determined by the behavior of the
source term $q(r,\om_q)$ as $r\to 1$. Since the wave function
$\psi^{(+)}_q(r)\sim (r-1)^{-2i\om_q}$ as $r\to 1$, the source term
(\ref{source_generic}) diverges as $(r-1)^{1-i\om_q}$ at the horizon.
Therefore $\zeta_q=-2i\om_q$ and the integral is, in general,
divergent.
The divergence can be regularized following the method proposed by
Detweiler and Szedenits~\cite{Detweiler:1979xr}. The idea is to add to
the integrand a total derivative which vanishes at the horizon:
\beq 
\label{regularize}
f(r)\equiv \frac{d}{dr}\left( \sum_{n=0}^N b_n
\frac{(r-1)^{\zeta_q+n+1}}{\zeta_q+n+1}e^{-(r-1)} \right)\,,
\eeq
where $N$ is greater than (or equal to) the largest integer in the
real part of $-2i\om_q$. For Schwarzschild infalls, the coefficients
$b_n$ in this expansion can be determined order-by-order in terms of
the $\xi_n$. The first few coefficients are listed in Appendix
\ref{app:Schwreg}, and the values of the ``excitation integrals''
$\widetilde{I}_q$ are listed in Table~\ref{tab:Iq}.

The values of the corresponding excitation coefficients
$\widetilde{C}_q=B_q^{(+)}\widetilde{I}_q$ are listed in
Table~\ref{tab:Cq}.  These tables were produced using a constant value
$N=2$ in Eq.~(\ref{regularize}), which is sufficient to regularize the
divergence for the first few overtones ($n=0\,,1\,,2\,,3$). We
verified that our results are insensitive to variations of $N$ within
at least six digits, as long as $N$ is large enough to eliminate the
divergence. 

\begin{figure*}[htb]
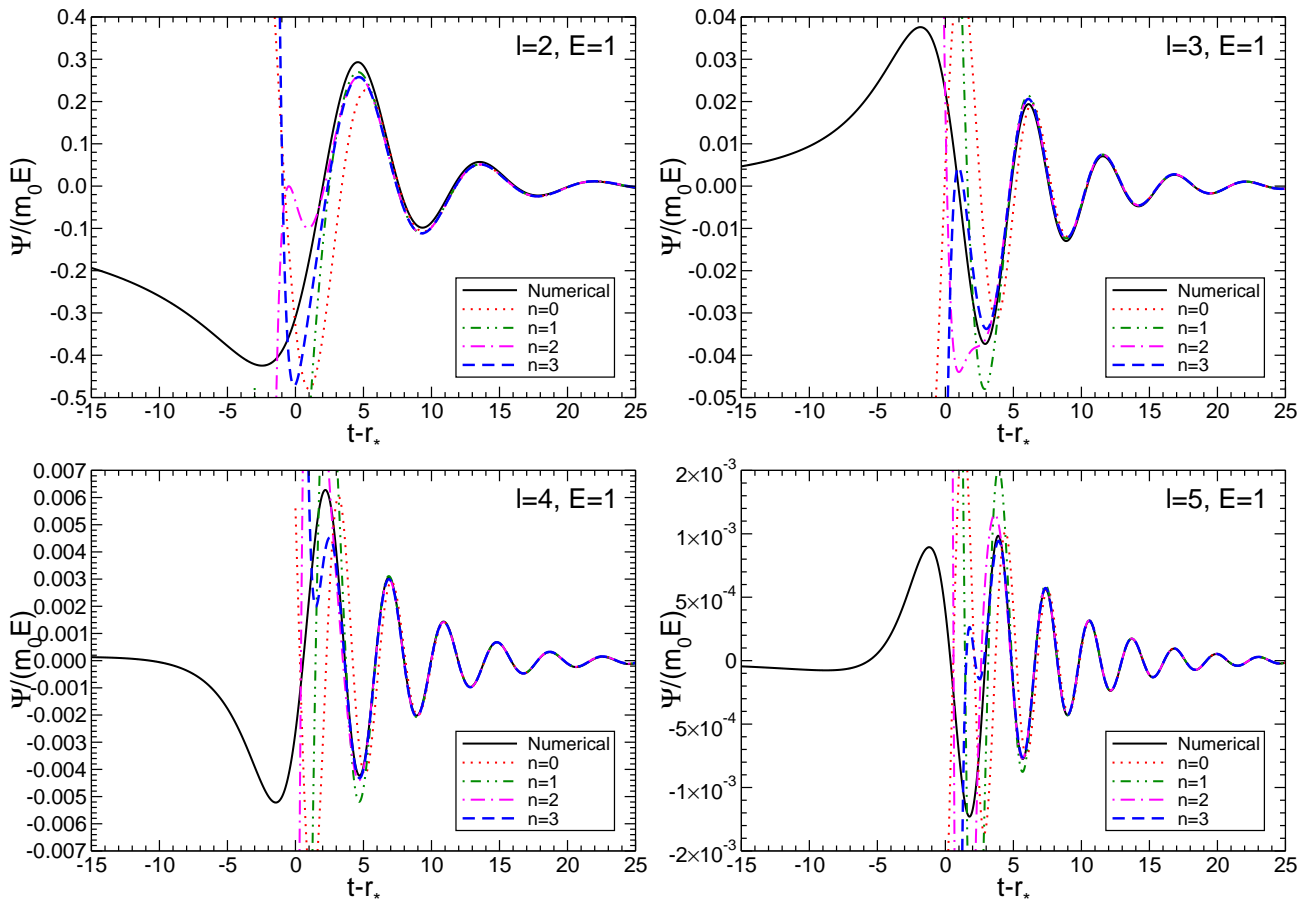

\begin{center}
\begin{tabular}{cc}
\epsfig{file=fig1a.eps,width=8.5cm,angle=0,clip=true}
&\epsfig{file=fig1b.eps,width=8.5cm,angle=0,clip=true}\\
\end{tabular}
\begin{tabular}{cc}
\epsfig{file=fig1c.eps,width=8.5cm,angle=0,clip=true}
&\epsfig{file=fig1d.eps,width=8.5cm,angle=0,clip=true}\\
\end{tabular}
\caption{\label{rest_l2345} Different multipolar components of the
  radiation ($l=2,3,4,5$) for an infall from rest. Solid black lines
  are obtained from a numerical solution of the perturbation equations
  in the Fourier domain \cite{Berti:2010ce,Sperhake:2011ik}, followed
  by an inverse Fourier transform. The other lines are obtained by
  summing an increasing numbers of overtones in the excitation
  coefficient calculation, as indicated in the legend. In this plot,
  as everywhere else in the paper, we use units $2M=1$.}
\end{center}
\end{figure*}

The tables show some interesting trends. For example, if we consider
infalls from rest ($E=1$) and a fixed multipolar index $l$, we see
that the real part of the excitation integral $\widetilde{I}_q$
increases as a function of the overtone index $n$. However this
increase is compensated by a comparable decrease in the imaginary part
of $\widetilde{I}_q$, so that $|\widetilde{I}_q|$ is roughly constant
as a function of $n$.

Figure \ref{rest_l2345} compares the excitation coefficient
calculation of Eq.~(\ref{semianal}) against numerical gravitational
waveforms for particles falling radially from rest. These waveforms
were computed using the frequency-domain codes described in
\cite{Berti:2010ce,Sperhake:2011ik}, and then Fourier-transformed back
in time. Each panel corresponds to a fixed multipole index
($l=2\,,3\,,4\,,5$), and different line styles correspond to ringdown
waveforms obtained summing a different number of overtones. This plot
generalizes and extends a similar comparison that can be found in
Fig.~10 of Leaver's original paper \cite{Leaver:1986gd}. Leaver found
a disagreement at the 10\% level, that he attributed to inaccuracies
in the Fourier transform of the numerical waveforms. We have similar
accuracy problems with the Fourier transform of our data (computing
Fourier amplitudes at low frequencies $\omega$ is time consuming,
because the computational domain must extend out to a radius $r\sim
1/\omega$), but the level of disagreement that we observe is smaller
than in Leaver's original analysis. Furthermore, the agreement between
our numerics and the excitation coefficient calculation gets better as
$l$ grows.  Figure \ref{rest_l2345} shows quite clearly that the
addition of higher overtones generally improves the agreement between
the excitation coefficient calculation and the full numerical waveform
at early times. However there is no analytical proof that the
expansion in terms of overtones should be convergent
\cite{Leaver:1986gd}, and indeed in a few isolated cases an expansion
including a large number of overtones can perform more poorly than a
similar expansion including a smaller number of overtones.

\begin{figure*}[htb]
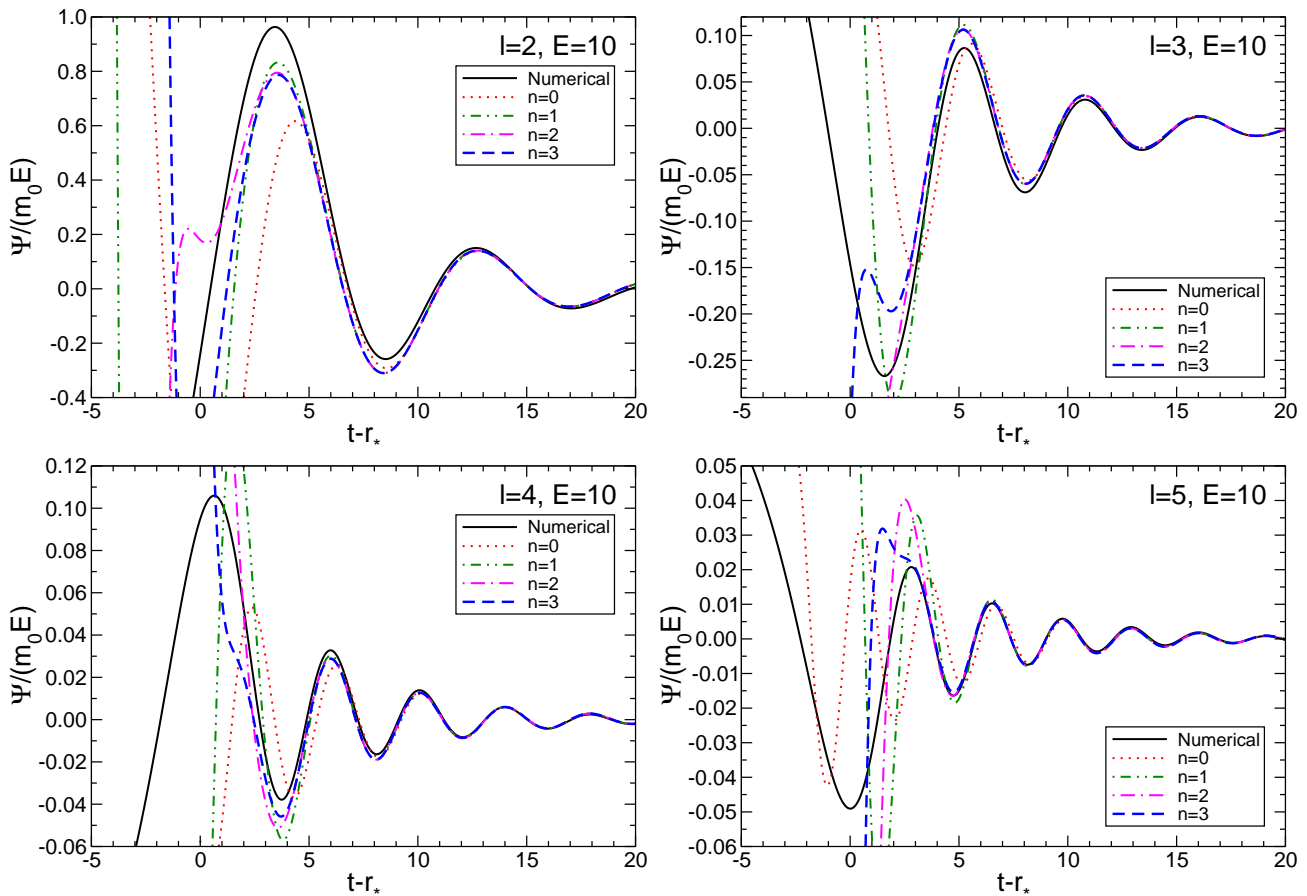

\begin{center}
\begin{tabular}{cc}
\epsfig{file=fig2a.eps,width=8.5cm,angle=0,clip=true}
&\epsfig{file=fig2b.eps,width=8.5cm,angle=0,clip=true}\\
\end{tabular}
\begin{tabular}{cc}
\epsfig{file=fig2c.eps,width=8.5cm,angle=0,clip=true}
&\epsfig{file=fig2d.eps,width=8.5cm,angle=0,clip=true}\\
\end{tabular}
\caption{\label{En10_l2345} Different multipolar components of the
  radiation ($l=2,3,4,5$) for an infall with initial energy
  $E=10$. Solid black lines are results from the numerical solution of
  the perturbation equations; the other lines are results obtained by
  summing different numbers of overtones. In this plot, as everywhere
  else in the paper, we use units $2M=1$.}
\end{center}
\end{figure*}

Figure \ref{En10_l2345} is similar to Figure~\ref{rest_l2345}, but it
refers to a relativistic infall with (normalized) particle energy
$E=10$. This figure shows that even by adding four overtones we don't
get excellent agreement at the ``absolute maximum'' of the numerical
waveform. Part of the reason is that we can only get accurate
numerical amplitudes at frequencies $M \om\gtrsim 10^{-3}$: to remove
``memory effects'' in the inverse Fourier transform, we extrapolate
our numerical calculations to obtain the Fourier-domain waveform
amplitude at frequencies $M\omega\lesssim 10^{-3}$. More importantly,
in ultrarelativistic infalls a larger fraction of the energy is
radiated during the infall (at low frequencies) than in the case of
infalls from rest. In other words, a larger fraction of the radiation
is produced {\it before} the beginning of the ringdown phase, and this
explains the larger disagreement between numerical waveforms and
``pure ringdown'' waveforms. As in the nonrelativistic case, we
observe that: (i) the ringdown waveform agrees better with the
numerical solution as $l$ grows; (ii) the addition of higher overtones
improves the agreement between the excitation coefficient expansion
and the numerical waveforms, but to a lesser extent, for the reasons
explained above.

\begin{figure*}[htb]
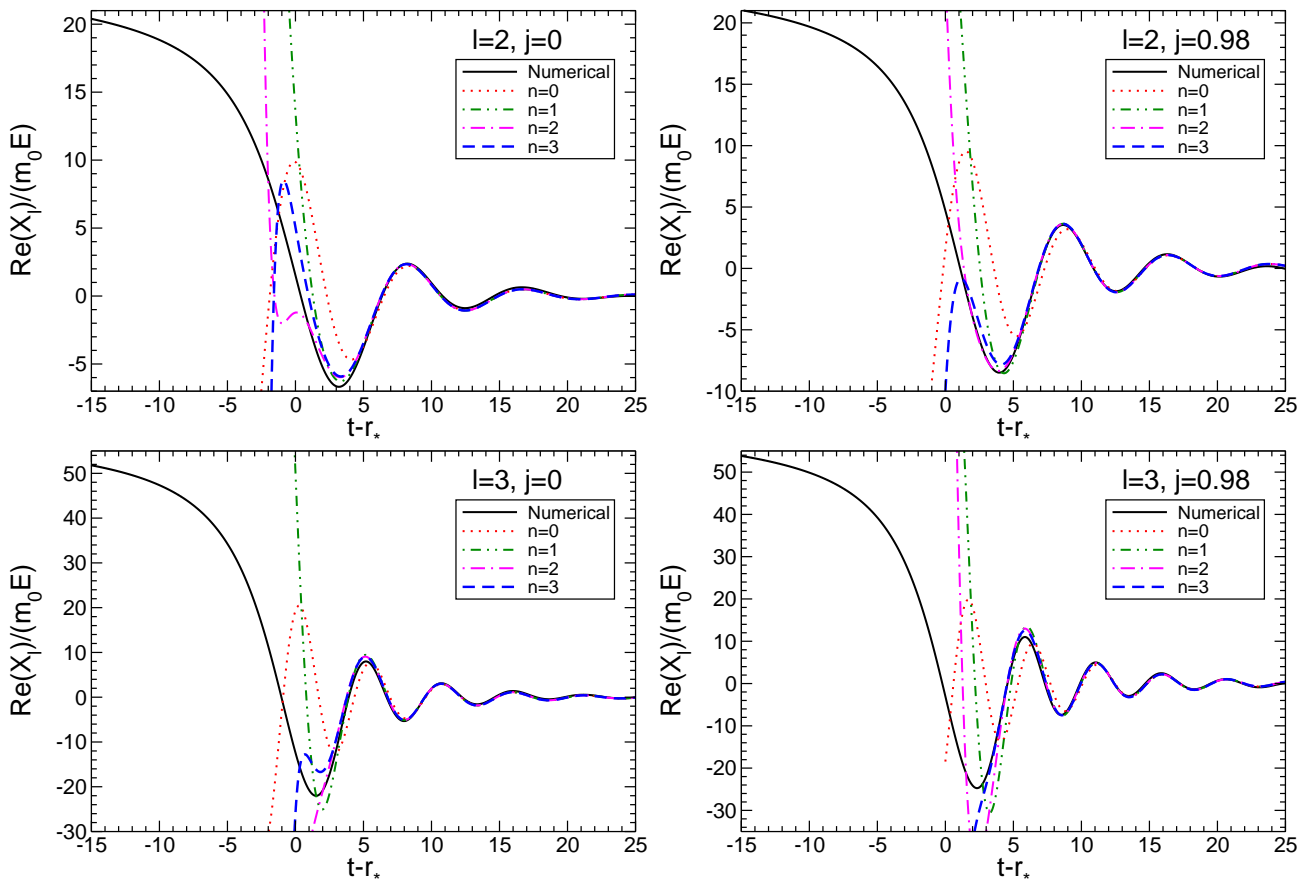

\begin{center}
\begin{tabular}{cc}
\epsfig{file=fig3a.eps,width=8.5cm,angle=0,clip=true}
&\epsfig{file=fig3b.eps,width=8.5cm,angle=0,clip=true}\\
\epsfig{file=fig3c.eps,width=8.5cm,angle=0,clip=true}
&\epsfig{file=fig3d.eps,width=8.5cm,angle=0,clip=true}\\
\end{tabular}
\caption{\label{Kerrplotsv2} Sasaki-Nakamura wave function for an
  ultrarelativistic infall along the symmetry axis of a Kerr BH. Solid
  black lines are results from the numerical solution of the
  perturbation equations; the other lines are results obtained by
  summing different numbers of overtones. The upper panels refer to
  $l=2$, the lower panels to $l=3$. The left panels corresponds to the
  Schwarzschild limit ($j=0$), and the right panels to a fast-spinning
  Kerr BH with $a=0.49$ ($j=0.98$). In this plot, as everywhere else
  in the paper, we use units $2M=1$.}
\end{center}
\end{figure*}

\section{Excitation factors and excitation coefficients for Kerr black holes}
\label{sec:Kerr}
In this section we extend our calculation to particles falling into
Kerr BHs.  For simplicity, we consider a particle falling
ultrarelativistically along the symmetry axis.  In this case the
source term of the Sasaki-Nakamura equation \eqref{eq:SN} simplifies
considerably \cite{Cardoso:2002yj,Cardoso:2002jr}:
\be {\cal S}=-\frac{m_0E C_l^a  \gamma \Delta}{2\omega^2 r^2 (r^2+a^2)^{3/2}}e^{-i\omega r_*}
\label{explicitS1} \,,
\ee
where
\be C_l^a=\lim_{\theta \to 0}\frac{8S_{l0\omega}(\theta,\phi)}{\sin^2\theta} \,,
\ee
and $\gamma$ was defined in \eqref{defgamma}. The constants $C_l^a$
were determined by solving the angular eigenvalue problem through a
continued fraction representation, and then plugging these eigenvalues
into the series solution providing the spheroidal wave functions
$S_{l0\omega}$ \cite{Berti:2005gp,Berti:2009kk}. The procedure to
determine the time-domain solution of the Sasaki-Nakamura wave
function $X$ is identical to that adopted for the Schwarzschild case:
i.e., first we solve the equations in the frequency domain, and then
we Fourier transform back in time, applying a low-frequency
extrapolation when this is necessary to remove memory effects.

Figure \ref{Kerrplotsv2} (which is similar to Figure \ref{rest_l2345})
compares the excitation coefficient calculation of
Eq.~(\ref{semianal}) -- where now $\Psi$ must be understood as the
Sasaki-Nakamura wave function -- against numerical gravitational
waveforms obtained in this way. As in the Schwarzschild case, the
integrand appearing in the calculation of the Kerr excitation factors
is, in general, divergent. The divergence can be regularized following
a procedure analogous to the Schwarzschild case (cf. Appendix
\ref{app:Kerrreg}).

Figure \ref{Kerrplotsv2} confirms our basic findings from the
nonrotating case: the convergence of the QNM expansion is not
necessarily monotonic, and the excitation coefficient expansion works
better for higher values of $l$. Notice that a relatively small number
of overtones is sufficient to reproduce the numerical waveform at
early times even when the spin of the Kerr BH is rather large
($j=0.98$), so that one may in principle expect that a larger number
of overtones would be necessary (see
e.g. \cite{Andersson:1999wj,Cardoso:2004hh,Berti:2005ys,Berti:2006hb,Yang:2012pj}). To
our knowledge, the calculation presented in this Section is the first
concrete proof that an excitation-coefficient expansion is applicable
and useful in the Kerr case: all calculations available in the
literature so far were specific to the Schwarzschild case (see
e.g. \cite{Hadar:2009ip,Hadar:2011vj}).

\section{Conclusions and outlook}
\label{sec:Conclusions}

In this paper we have implemented a new method, based on the MST
formalism, to compute the excitation factors $B_q$ for Kerr QNMs.
This method is simpler and more accurate than the method used by two
of us in \cite{Berti:2006wq}, allowing us to extend the calculation to
higher angular multipoles $l$ and to higher overtone numbers $n$.
Tables of the excitation factors $B_q$ in the Teukolsky and
Sasaki-Nakamura formalisms will be made publicly available online
\cite{rdweb}, in the hope to stimulate further research in this field.

As a test of the method, we have computed the QNM excitation
coefficients for the classic problem of particles falling radially
into the BH. We have compared the excitation coefficient expansion
against numerical results for: (i) particles falling from rest ($E=1$)
into a Schwarzschild BH, (ii) large-energy particles ($E=10$) falling
into a Schwarzschild BH, and (iii) ultrarelativistic particles falling
into a Kerr BH along the symmetry axis. In all cases we found
excellent agreement, validating the usefulness of excitation
coefficient calculations in the analytical modeling of the ringdown
phase.

In order of increasing complexity, extensions of this work could
consider (i) particles falling with arbitrary energy along the
$z$-axis of a Kerr BH, (ii) particles with arbitrary energies plunging
into Kerr BHs along equatorial orbits, (iii) generic orbits in
Schwarzschild or Kerr, and (iv) possible applications of these
calculations to the construction of semianalytical waveform templates
for comparable-mass mergers. We believe that these extensions are
crucial for a better understanding of the ringdown phase and (more
ambitiously) for the construction of gravitational-wave detection
templates for comparable-mass BH binaries.

\section{Acknowledgments}
We are grateful to Sam Dolan for correspondence on the calculation of
the excitation factors. E.B. and Z.Z.'s research was supported by NSF
Grant No. PHY-0900735 and NSF CAREER Grant No. PHY-1055103.
V.C. acknowledges partial financial support provided under the
European Union's FP7 ERC Starting Grant ``The dynamics of black holes:
testing the limits of Einstein's theory'' grant agreement no.
DyBHo--256667, the NRHEP 295189 FP7-PEOPLE-2011-IRSES Grant, and
FCT-Portugal through project CERN/FP/123593/2011.  Research at
Perimeter Institute is supported by the Government of Canada through
Industry Canada and by the Province of Ontario through the Ministry of
Economic Development and Innovation.

\appendix

\section{Regularization coefficients}
\label{app:reg}

\subsection{The Schwarzschild case}
\label{app:Schwreg}

For reference, in this Appendix we list the first few regularization
coefficients $b_n$ defined in Eq.~(\ref{regularize}):
\beq  
b_0&=&\xi_0,\\
b_1&=&\xi_1+\frac{2-2i\om_q}{1-2i\om_q}~b_0,\nn\\
b_2&=&\xi_2+\frac{2i\om_q-3}{2(1-2i\om_q)}~b_0+\frac{3-2i\om_q}{2(1-i\om_q)}~b_1,\nn\\
b_3&=&\xi_3
+\frac{2-i\om_q}{3(1-2i\om_q)}~b_0+\frac{i\om_q-2}{2(1-i\om_q)}~b_1+\frac{2(2-i\om_q)}{3-2i\om_q}~b_2
,\nn\\
b_4&=&\xi_4+
\frac{2i\om_q-5}{24(1-2i\om_q)}~b_0+\frac{5-2i\om_q}{12(1-i\om_q)}~b_1\nn\\
&+&\frac{2i\om_q-5}{2(3-2i\om_q)}~b_2
+\frac{5-2i\om_q}{2(2-i\om_q)}~b_3  \,.\nn
\eeq

\subsection{The Kerr case}
\label{app:Kerrreg}

The regularization coefficients for the Kerr case are much more
lengthy than in the nonrotating case, but their calculation is
straightforward. Here we list for reference the first two
coefficients:
\begin{widetext}
\beq
b_0 &=& \frac{1}{A_{\rm out}~\omega_q^2r_+^2}(r_+-r_-)^{\left(\frac{2i\omega_q r_+}{r_+-r_-}-1\right)}
(ir_{+}-ir_{-}+\omega_q\,r_{+})(2ir_{+}-i+2\omega_q r_+)\,,\\
b_1&=&2b_0\frac{r_{-}-r_{+}+i\omega_q r_+}{r_{-}-r_{+}+2i\omega_{q}r_+}+\frac{1}{2A_{\rm out}^{\text{SN}}\omega_{q}^2}
\frac{1}{r_{+}^3}(r_{+}-r_{-})^{\frac{3r_{+}-3r_{-}-2i\omega_{q}r_+}{r_{-}-r_{+}}}\notag\\
&\times&\Bigg\{r_{-}^4(8+4ir_{+}\omega_q)+r_{-}^3\bigg[4+r_+(-36+\lambda+12i\omega_q)-2r_+^2\omega_q(5i+2\omega_{q})\bigg]\notag\\
&+&r_{-}^2r_{+}\bigg[-11+4i\omega_{q}+6r_{+}^2\omega_q (i+2\omega_q)+3a_{r1}-r_{+}(-58+3\lambda+38i\omega_{q}+4\omega_q^2
+6a_{r1}-6i\omega_{q}a_{r1})\bigg]\notag\\
&+&r_{+}^3\bigg[4\omega_q^2+2r_+^2\omega_q(-i+2\omega_q)-2i\omega_q(-5+a_ {r1})+3(-1+a_ {r1})-r_{+}\Big(\lambda+2(1-i\omega_q)
\big(-5+4\omega_q^2-2i\omega_q(-4+a_{r1})+3a_ {r1}\big)\Big)\bigg]\notag\\
&+&r_{-}r_+^2\bigg[2r_{+}^2(i-6\omega_q)\omega_q+2\big(5+i\omega_q(-7+a_ {r1})-3a_ {r1}\big)+r_+\Big(3\lambda-4\big(10+\omega_q^2(-5+a_ {r1})-
3a_{r1}+i\omega_q(-13+4a_{r1})\big)\Big)\bigg]\Bigg\}\,,\notag
\eeq
\end{widetext}
where
\be
\sigma_+=\frac{\omega_q r_+ -a m}{r_+-r_-}\,,
\ee
the amplitude $A_{\rm out}^{\rm SN}$ is related to the Teukolsky
amplitude $A_{\rm out}^{\rm T}=\sum_{n=0}^\infty a_{rn}$ via
Eq.~(\ref{relasympt}) and $\lambda$ is related to the separation
constant $A_{lm}$ through relation (\ref{eq:lambda}).  The
coefficients $\{a_{rn}\}$, $n=0,\,1,\,2...$ (with $a_{r0}=1$) are
defined via the homogeneous solution $R_{r_{+}}$ of the Teukolsky
equation
\begin{widetext}
\beq  
R_{r_{+}}=e^{i\omega_q
r}(r-r_-)^{-1-s+i\omega_q+i\sigma_+}(r-r_+)^{-s-i\sigma_+}
\sum_{n=0}^\infty a_{rn}\left(\frac{r-r_+}{r-r_-}\right)^n\,,
\eeq
\end{widetext}
and can be obtained by plugging this decomposition in the Teukolsky
equation (\ref{r1}).

The Sasaki-Nakamura wave function $X$ is related to $R_{r_{+}}$ by
Eq.~(\ref{XToR}). What we plot in Figure \ref{Kerrplotsv2} is actually
the normalized Sasaki-Nakamura wave form $X_q^{\text{SN}}=(X
e^{-i\omega_qr_*})/A_{\rm out}$,
whose excitation coefficients are given by
\begin{widetext}
\beq  
C_q =
-\gamma_{\infty}\int_{r_+}^{\infty}dr\left[\frac{X_q^{\text{SN}}}{2\omega_q^2r^2\sqrt{(r-r_+)(r-r_-)+r}}
-\sum_{k=0}^\infty\left(e^{-r+r_+}(r-r_+)^{k+\frac{2i\omega_qr_+}{r_--r_+}}b_k
-\frac{e^{-r+r_+}(r-r_+)^{1+k+\frac{2i\omega_qr_+}{r_--r_+}}}{1+k+\frac{2i\omega_qr_+}{r_--r_+}}b_k\right)\right]\,.\nn
\eeq
\end{widetext}


\begin{thebibliography}{57}%
\makeatletter
\providecommand \@ifxundefined [1]{%
 \@ifx{#1\undefined}
}%
\providecommand \@ifnum [1]{%
 \ifnum #1\expandafter \@firstoftwo
 \else \expandafter \@secondoftwo
 \fi
}%
\providecommand \@ifx [1]{%
 \ifx #1\expandafter \@firstoftwo
 \else \expandafter \@secondoftwo
 \fi
}%
\providecommand \natexlab [1]{#1}%
\providecommand \enquote  [1]{``#1''}%
\providecommand \bibnamefont  [1]{#1}%
\providecommand \bibfnamefont [1]{#1}%
\providecommand \citenamefont [1]{#1}%
\providecommand \href@noop [0]{\@secondoftwo}%
\providecommand \href [0]{\begingroup \@sanitize@url \@href}%
\providecommand \@href[1]{\@@startlink{#1}\@@href}%
\providecommand \@@href[1]{\endgroup#1\@@endlink}%
\providecommand \@sanitize@url [0]{\catcode `\\12\catcode `\$12\catcode
  `\&12\catcode `\#12\catcode `\^12\catcode `\_12\catcode `\%12\relax}%
\providecommand \@@startlink[1]{}%
\providecommand \@@endlink[0]{}%
\providecommand \url  [0]{\begingroup\@sanitize@url \@url }%
\providecommand \@url [1]{\endgroup\@href {#1}{\urlprefix }}%
\providecommand \urlprefix  [0]{URL }%
\providecommand \Eprint [0]{\href }%
\providecommand \doibase [0]{http://dx.doi.org/}%
\providecommand \selectlanguage [0]{\@gobble}%
\providecommand \bibinfo  [0]{\@secondoftwo}%
\providecommand \bibfield  [0]{\@secondoftwo}%
\providecommand \translation [1]{[#1]}%
\providecommand \BibitemOpen [0]{}%
\providecommand \bibitemStop [0]{}%
\providecommand \bibitemNoStop [0]{.\EOS\space}%
\providecommand \EOS [0]{\spacefactor3000\relax}%
\providecommand \BibitemShut  [1]{\csname bibitem#1\endcsname}%
\let\auto@bib@innerbib\@empty
\bibitem [{\citenamefont {Leaver}(1986)}]{Leaver:1986gd}%
  \BibitemOpen
  \bibfield  {author} {\bibinfo {author} {\bibfnamefont {E.~W.}\ \bibnamefont
  {Leaver}},\ }\href {\doibase 10.1103/PhysRevD.34.384} {\bibfield  {journal}
  {\bibinfo  {journal} {Phys. Rev.}\ }\textbf {\bibinfo {volume} {D34}},\
  \bibinfo {pages} {384} (\bibinfo {year} {1986})}\BibitemShut {NoStop}%
\bibitem [{\citenamefont {Berti}\ and\ \citenamefont
  {Cardoso}(2006)}]{Berti:2006wq}%
  \BibitemOpen
  \bibfield  {author} {\bibinfo {author} {\bibfnamefont {E.}~\bibnamefont
  {Berti}}\ and\ \bibinfo {author} {\bibfnamefont {V.}~\bibnamefont
  {Cardoso}},\ }\href {\doibase 10.1103/PhysRevD.74.104020} {\bibfield
  {journal} {\bibinfo  {journal} {Phys. Rev.}\ }\textbf {\bibinfo {volume}
  {D74}},\ \bibinfo {pages} {104020} (\bibinfo {year} {2006})},\ \Eprint
  {http://arxiv.org/abs/gr-qc/0605118} {arXiv:gr-qc/0605118} \BibitemShut
  {NoStop}%
\bibitem [{\citenamefont {Kokkotas}\ and\ \citenamefont
  {Schmidt}(1999)}]{Kokkotas:1999bd}%
  \BibitemOpen
  \bibfield  {author} {\bibinfo {author} {\bibfnamefont {K.~D.}\ \bibnamefont
  {Kokkotas}}\ and\ \bibinfo {author} {\bibfnamefont {B.~G.}\ \bibnamefont
  {Schmidt}},\ }\href {http://www.livingreviews.org/lrr-1999-2} {\bibfield
  {journal} {\bibinfo  {journal} {Living Rev. Relativity}\ }\textbf {\bibinfo
  {volume} {2}} (\bibinfo {year} {1999})},\ \Eprint
  {http://arxiv.org/abs/gr-qc/9909058} {arXiv:gr-qc/9909058} \BibitemShut
  {NoStop}%
\bibitem [{\citenamefont {Nollert}(1999)}]{Nollert:1999ji}%
  \BibitemOpen
  \bibfield  {author} {\bibinfo {author} {\bibfnamefont {H.-P.}\ \bibnamefont
  {Nollert}},\ }\href {\doibase 10.1088/0264-9381/16/12/201} {\bibfield
  {journal} {\bibinfo  {journal} {Class. Quantum Grav.}\ }\textbf {\bibinfo
  {volume} {16}},\ \bibinfo {pages} {R159} (\bibinfo {year}
  {1999})}\BibitemShut {NoStop}%
\bibitem [{\citenamefont {Berti}\ \emph {et~al.}(2009)\citenamefont {Berti},
  \citenamefont {Cardoso},\ and\ \citenamefont {Starinets}}]{Berti:2009kk}%
  \BibitemOpen
  \bibfield  {author} {\bibinfo {author} {\bibfnamefont {E.}~\bibnamefont
  {Berti}}, \bibinfo {author} {\bibfnamefont {V.}~\bibnamefont {Cardoso}}, \
  and\ \bibinfo {author} {\bibfnamefont {A.~O.}\ \bibnamefont {Starinets}},\
  }\href {\doibase 10.1088/0264-9381/26/16/163001} {\bibfield  {journal}
  {\bibinfo  {journal} {Class. Quantum Grav.}\ }\textbf {\bibinfo {volume}
  {26}},\ \bibinfo {pages} {163001} (\bibinfo {year} {2009})},\ \Eprint
  {http://arxiv.org/abs/0905.2975} {arXiv:0905.2975 [gr-qc]} \BibitemShut
  {NoStop}%
\bibitem [{\citenamefont {Pretorius}(2007)}]{Pretorius:2007nq}%
  \BibitemOpen
  \bibfield  {author} {\bibinfo {author} {\bibfnamefont {F.}~\bibnamefont
  {Pretorius}},\ }\enquote {\bibinfo {title} {{Binary Black Hole
  Coalescence}},}\ in\ \href@noop {} {\emph {\bibinfo {booktitle} {Physics of
  relativistic objects in compact binaries: from birth to coalescence
  (Astrophysics and Space Science Library, Vol. 359)}}}\ (\bibinfo  {publisher}
  {Springer},\ \bibinfo {address} {New York},\ \bibinfo {year} {2007})\
  Chap.~\bibinfo {chapter} {9},\ \Eprint {http://arxiv.org/abs/0710.1338}
  {arXiv:0710.1338 [gr-qc]} \BibitemShut {NoStop}%
\bibitem [{\citenamefont {Sperhake}\ \emph
  {et~al.}(2011{\natexlab{a}})\citenamefont {Sperhake}, \citenamefont {Berti},\
  and\ \citenamefont {Cardoso}}]{Sperhake:2011xk}%
  \BibitemOpen
  \bibfield  {author} {\bibinfo {author} {\bibfnamefont {U.}~\bibnamefont
  {Sperhake}}, \bibinfo {author} {\bibfnamefont {E.}~\bibnamefont {Berti}}, \
  and\ \bibinfo {author} {\bibfnamefont {V.}~\bibnamefont {Cardoso}},\
  }\href@noop {} {\  (\bibinfo {year} {2011}{\natexlab{a}})},\ \Eprint
  {http://arxiv.org/abs/1107.2819} {arXiv:1107.2819 [gr-qc]} \BibitemShut
  {NoStop}%
\bibitem [{\citenamefont {Pfeiffer}(2012)}]{Pfeiffer:2012pc}%
  \BibitemOpen
  \bibfield  {author} {\bibinfo {author} {\bibfnamefont {H.~P.}\ \bibnamefont
  {Pfeiffer}},\ }\href {\doibase 10.1088/0264-9381/29/12/124004} {\bibfield
  {journal} {\bibinfo  {journal} {Class.Quant.Grav.}\ }\textbf {\bibinfo
  {volume} {29}},\ \bibinfo {pages} {124004} (\bibinfo {year} {2012})},\
  \Eprint {http://arxiv.org/abs/1203.5166} {arXiv:1203.5166 [gr-qc]}
  \BibitemShut {NoStop}%
\bibitem [{\citenamefont {Sperhake}(2013)}]{Sperhake:2013qa}%
  \BibitemOpen
  \bibfield  {author} {\bibinfo {author} {\bibfnamefont {U.}~\bibnamefont
  {Sperhake}},\ }\href@noop {} {\  (\bibinfo {year} {2013})},\ \Eprint
  {http://arxiv.org/abs/1301.3772} {arXiv:1301.3772 [gr-qc]} \BibitemShut
  {NoStop}%
\bibitem [{\citenamefont {Gonz{\'a}lez}\ \emph {et~al.}(2009)\citenamefont
  {Gonz{\'a}lez}, \citenamefont {Sperhake},\ and\ \citenamefont
  {Br{\"u}gmann}}]{Gonzalez:2008bi}%
  \BibitemOpen
  \bibfield  {author} {\bibinfo {author} {\bibfnamefont {J.~A.}\ \bibnamefont
  {Gonz{\'a}lez}}, \bibinfo {author} {\bibfnamefont {U.}~\bibnamefont
  {Sperhake}}, \ and\ \bibinfo {author} {\bibfnamefont {B.}~\bibnamefont
  {Br{\"u}gmann}},\ }\href {\doibase 10.1103/PhysRevD.79.124006} {\bibfield
  {journal} {\bibinfo  {journal} {Phys. Rev.}\ }\textbf {\bibinfo {volume}
  {D79}},\ \bibinfo {pages} {124006} (\bibinfo {year} {2009})},\ \Eprint
  {http://arxiv.org/abs/0811.3952} {arXiv:0811.3952 [gr-qc]} \BibitemShut
  {NoStop}%
\bibitem [{\citenamefont {Lousto}\ \emph {et~al.}(2010)\citenamefont {Lousto},
  \citenamefont {Nakano}, \citenamefont {Zlochower},\ and\ \citenamefont
  {Campanelli}}]{Lousto:2010qx}%
  \BibitemOpen
  \bibfield  {author} {\bibinfo {author} {\bibfnamefont {C.~O.}\ \bibnamefont
  {Lousto}}, \bibinfo {author} {\bibfnamefont {H.}~\bibnamefont {Nakano}},
  \bibinfo {author} {\bibfnamefont {Y.}~\bibnamefont {Zlochower}}, \ and\
  \bibinfo {author} {\bibfnamefont {M.}~\bibnamefont {Campanelli}},\ }\href
  {\doibase 10.1103/PhysRevD.82.104057} {\bibfield  {journal} {\bibinfo
  {journal} {Phys. Rev.}\ }\textbf {\bibinfo {volume} {D82}},\ \bibinfo {pages}
  {104057} (\bibinfo {year} {2010})},\ \Eprint {http://arxiv.org/abs/1008.4360}
  {arXiv:1008.4360 [gr-qc]} \BibitemShut {NoStop}%
\bibitem [{\citenamefont {Nakano}\ \emph {et~al.}(2011)\citenamefont {Nakano},
  \citenamefont {Zlochower}, \citenamefont {Lousto},\ and\ \citenamefont
  {Campanelli}}]{Nakano:2011pb}%
  \BibitemOpen
  \bibfield  {author} {\bibinfo {author} {\bibfnamefont {H.}~\bibnamefont
  {Nakano}}, \bibinfo {author} {\bibfnamefont {Y.}~\bibnamefont {Zlochower}},
  \bibinfo {author} {\bibfnamefont {C.~O.}\ \bibnamefont {Lousto}}, \ and\
  \bibinfo {author} {\bibfnamefont {M.}~\bibnamefont {Campanelli}},\ }\href
  {\doibase 10.1103/PhysRevD.84.124006} {\bibfield  {journal} {\bibinfo
  {journal} {Phys. Rev.}\ }\textbf {\bibinfo {volume} {D84}},\ \bibinfo {pages}
  {124006} (\bibinfo {year} {2011})},\ \Eprint {http://arxiv.org/abs/1108.4421}
  {arXiv:1108.4421 [gr-qc]} \BibitemShut {NoStop}%
\bibitem [{\citenamefont {Sperhake}\ \emph
  {et~al.}(2011{\natexlab{b}})\citenamefont {Sperhake}, \citenamefont
  {Cardoso}, \citenamefont {Ott}, \citenamefont {Schnetter},\ and\
  \citenamefont {Witek}}]{Sperhake:2011ik}%
  \BibitemOpen
  \bibfield  {author} {\bibinfo {author} {\bibfnamefont {U.}~\bibnamefont
  {Sperhake}}, \bibinfo {author} {\bibfnamefont {V.}~\bibnamefont {Cardoso}},
  \bibinfo {author} {\bibfnamefont {C.~D.}\ \bibnamefont {Ott}}, \bibinfo
  {author} {\bibfnamefont {E.}~\bibnamefont {Schnetter}}, \ and\ \bibinfo
  {author} {\bibfnamefont {H.}~\bibnamefont {Witek}},\ }\href {\doibase
  10.1103/PhysRevD.84.084038} {\bibfield  {journal} {\bibinfo  {journal}
  {Phys.Rev.}\ }\textbf {\bibinfo {volume} {D84}},\ \bibinfo {pages} {084038}
  (\bibinfo {year} {2011}{\natexlab{b}})},\ \Eprint
  {http://arxiv.org/abs/1105.5391} {arXiv:1105.5391 [gr-qc]} \BibitemShut
  {NoStop}%
\bibitem [{\citenamefont {East}\ and\ \citenamefont
  {Pretorius}(2013)}]{East:2013iwa}%
  \BibitemOpen
  \bibfield  {author} {\bibinfo {author} {\bibfnamefont {W.~E.}\ \bibnamefont
  {East}}\ and\ \bibinfo {author} {\bibfnamefont {F.}~\bibnamefont
  {Pretorius}},\ }\href@noop {} {\  (\bibinfo {year} {2013})},\ \Eprint
  {http://arxiv.org/abs/1303.1540} {arXiv:1303.1540 [gr-qc]} \BibitemShut
  {NoStop}%
\bibitem [{\citenamefont {Lousto}\ \emph {et~al.}(2012)\citenamefont {Lousto},
  \citenamefont {Nakano}, \citenamefont {Zlochower}, \citenamefont {Mundim},\
  and\ \citenamefont {Campanelli}}]{Lousto:2012es}%
  \BibitemOpen
  \bibfield  {author} {\bibinfo {author} {\bibfnamefont {C.~O.}\ \bibnamefont
  {Lousto}}, \bibinfo {author} {\bibfnamefont {H.}~\bibnamefont {Nakano}},
  \bibinfo {author} {\bibfnamefont {Y.}~\bibnamefont {Zlochower}}, \bibinfo
  {author} {\bibfnamefont {B.~C.}\ \bibnamefont {Mundim}}, \ and\ \bibinfo
  {author} {\bibfnamefont {M.}~\bibnamefont {Campanelli}},\ }\href {\doibase
  10.1103/PhysRevD.85.124013} {\bibfield  {journal} {\bibinfo  {journal}
  {Phys.Rev.}\ }\textbf {\bibinfo {volume} {D85}},\ \bibinfo {pages} {124013}
  (\bibinfo {year} {2012})},\ \Eprint {http://arxiv.org/abs/1203.3223}
  {arXiv:1203.3223 [gr-qc]} \BibitemShut {NoStop}%
\bibitem [{\citenamefont {Lovelace}\ \emph {et~al.}(2008)\citenamefont
  {Lovelace}, \citenamefont {Owen}, \citenamefont {Pfeiffer},\ and\
  \citenamefont {Chu}}]{Lovelace:2008tw}%
  \BibitemOpen
  \bibfield  {author} {\bibinfo {author} {\bibfnamefont {G.}~\bibnamefont
  {Lovelace}}, \bibinfo {author} {\bibfnamefont {R.}~\bibnamefont {Owen}},
  \bibinfo {author} {\bibfnamefont {H.~P.}\ \bibnamefont {Pfeiffer}}, \ and\
  \bibinfo {author} {\bibfnamefont {T.}~\bibnamefont {Chu}},\ }\href {\doibase
  10.1103/PhysRevD.78.084017} {\bibfield  {journal} {\bibinfo  {journal}
  {Phys.Rev.}\ }\textbf {\bibinfo {volume} {D78}},\ \bibinfo {pages} {084017}
  (\bibinfo {year} {2008})},\ \Eprint {http://arxiv.org/abs/0805.4192}
  {arXiv:0805.4192 [gr-qc]} \BibitemShut {NoStop}%
\bibitem [{\citenamefont {Lovelace}\ \emph {et~al.}(2011)\citenamefont
  {Lovelace}, \citenamefont {Scheel},\ and\ \citenamefont
  {Szilagyi}}]{Lovelace:2010ne}%
  \BibitemOpen
  \bibfield  {author} {\bibinfo {author} {\bibfnamefont {G.}~\bibnamefont
  {Lovelace}}, \bibinfo {author} {\bibfnamefont {M.}~\bibnamefont {Scheel}}, \
  and\ \bibinfo {author} {\bibfnamefont {B.}~\bibnamefont {Szilagyi}},\ }\href
  {\doibase 10.1103/PhysRevD.83.024010} {\bibfield  {journal} {\bibinfo
  {journal} {Phys.Rev.}\ }\textbf {\bibinfo {volume} {D83}},\ \bibinfo {pages}
  {024010} (\bibinfo {year} {2011})},\ \Eprint {http://arxiv.org/abs/1010.2777}
  {arXiv:1010.2777 [gr-qc]} \BibitemShut {NoStop}%
\bibitem [{\citenamefont {Lovelace}\ \emph {et~al.}(2012)\citenamefont
  {Lovelace}, \citenamefont {Boyle}, \citenamefont {Scheel},\ and\
  \citenamefont {Szilagyi}}]{Lovelace:2011nu}%
  \BibitemOpen
  \bibfield  {author} {\bibinfo {author} {\bibfnamefont {G.}~\bibnamefont
  {Lovelace}}, \bibinfo {author} {\bibfnamefont {M.}~\bibnamefont {Boyle}},
  \bibinfo {author} {\bibfnamefont {M.~A.}\ \bibnamefont {Scheel}}, \ and\
  \bibinfo {author} {\bibfnamefont {B.}~\bibnamefont {Szilagyi}},\ }\href
  {\doibase 10.1088/0264-9381/29/4/045003} {\bibfield  {journal} {\bibinfo
  {journal} {Class. Quantum Grav.}\ }\textbf {\bibinfo {volume} {29}},\
  \bibinfo {pages} {045003} (\bibinfo {year} {2012})},\ \Eprint
  {http://arxiv.org/abs/1110.2229} {arXiv:1110.2229 [gr-qc]} \BibitemShut
  {NoStop}%
\bibitem [{\citenamefont {Taracchini}\ \emph {et~al.}(2012)\citenamefont
  {Taracchini}, \citenamefont {Pan}, \citenamefont {Buonanno}, \citenamefont
  {Barausse}, \citenamefont {Boyle} \emph {et~al.}}]{Taracchini:2012ig}%
  \BibitemOpen
  \bibfield  {author} {\bibinfo {author} {\bibfnamefont {A.}~\bibnamefont
  {Taracchini}}, \bibinfo {author} {\bibfnamefont {Y.}~\bibnamefont {Pan}},
  \bibinfo {author} {\bibfnamefont {A.}~\bibnamefont {Buonanno}}, \bibinfo
  {author} {\bibfnamefont {E.}~\bibnamefont {Barausse}}, \bibinfo {author}
  {\bibfnamefont {M.}~\bibnamefont {Boyle}},  \emph {et~al.},\ }\href {\doibase
  10.1103/PhysRevD.86.024011} {\bibfield  {journal} {\bibinfo  {journal}
  {Phys.Rev.}\ }\textbf {\bibinfo {volume} {D86}},\ \bibinfo {pages} {024011}
  (\bibinfo {year} {2012})},\ \Eprint {http://arxiv.org/abs/1202.0790}
  {arXiv:1202.0790 [gr-qc]} \BibitemShut {NoStop}%
\bibitem [{\citenamefont {Berti}\ \emph {et~al.}(2004)\citenamefont {Berti},
  \citenamefont {Cavagli{\`a}},\ and\ \citenamefont
  {Gualtieri}}]{Berti:2003si}%
  \BibitemOpen
  \bibfield  {author} {\bibinfo {author} {\bibfnamefont {E.}~\bibnamefont
  {Berti}}, \bibinfo {author} {\bibfnamefont {M.}~\bibnamefont {Cavagli{\`a}}},
  \ and\ \bibinfo {author} {\bibfnamefont {L.}~\bibnamefont {Gualtieri}},\
  }\href {\doibase 10.1103/PhysRevD.69.124011} {\bibfield  {journal} {\bibinfo
  {journal} {Phys. Rev.}\ }\textbf {\bibinfo {volume} {D69}},\ \bibinfo {pages}
  {124011} (\bibinfo {year} {2004})},\ \Eprint
  {http://arxiv.org/abs/hep-th/0309203} {arXiv:hep-th/0309203} \BibitemShut
  {NoStop}%
\bibitem [{\citenamefont {Berti}\ \emph {et~al.}(2011)\citenamefont {Berti},
  \citenamefont {Cardoso},\ and\ \citenamefont {Kipapa}}]{Berti:2010gx}%
  \BibitemOpen
  \bibfield  {author} {\bibinfo {author} {\bibfnamefont {E.}~\bibnamefont
  {Berti}}, \bibinfo {author} {\bibfnamefont {V.}~\bibnamefont {Cardoso}}, \
  and\ \bibinfo {author} {\bibfnamefont {B.}~\bibnamefont {Kipapa}},\ }\href
  {\doibase 10.1103/PhysRevD.83.084018} {\bibfield  {journal} {\bibinfo
  {journal} {Phys. Rev.}\ }\textbf {\bibinfo {volume} {D83}},\ \bibinfo {pages}
  {084018} (\bibinfo {year} {2011})},\ \Eprint {http://arxiv.org/abs/1010.3874}
  {arXiv:1010.3874 [gr-qc]} \BibitemShut {NoStop}%
\bibitem [{\citenamefont {Witek}\ \emph {et~al.}(2011)\citenamefont {Witek},
  \citenamefont {Cardoso}, \citenamefont {Gualtieri}, \citenamefont {Herdeiro},
  \citenamefont {Sperhake},\ and\ \citenamefont {Zilh{\~a}o}}]{Witek:2010az}%
  \BibitemOpen
  \bibfield  {author} {\bibinfo {author} {\bibfnamefont {H.}~\bibnamefont
  {Witek}}, \bibinfo {author} {\bibfnamefont {V.}~\bibnamefont {Cardoso}},
  \bibinfo {author} {\bibfnamefont {L.}~\bibnamefont {Gualtieri}}, \bibinfo
  {author} {\bibfnamefont {C.}~\bibnamefont {Herdeiro}}, \bibinfo {author}
  {\bibfnamefont {U.}~\bibnamefont {Sperhake}}, \ and\ \bibinfo {author}
  {\bibfnamefont {M.}~\bibnamefont {Zilh{\~a}o}},\ }\href {\doibase
  10.1103/PhysRevD.83.044017} {\bibfield  {journal} {\bibinfo  {journal} {Phys.
  Rev.}\ }\textbf {\bibinfo {volume} {D83}},\ \bibinfo {pages} {044017}
  (\bibinfo {year} {2011})},\ \Eprint {http://arxiv.org/abs/1011.0742}
  {arXiv:1011.0742 [gr-qc]} \BibitemShut {NoStop}%
\bibitem [{\citenamefont {Yoshino}\ and\ \citenamefont
  {Shibata}(2011)}]{Yoshino:2011zz}%
  \BibitemOpen
  \bibfield  {author} {\bibinfo {author} {\bibfnamefont {H.~M.~S.}\
  \bibnamefont {Yoshino}}\ and\ \bibinfo {author} {\bibfnamefont
  {M.}~\bibnamefont {Shibata}},\ }\href {\doibase 10.1143/PTPS.189.269}
  {\bibfield  {journal} {\bibinfo  {journal} {Prog.Theor.Phys.Suppl.}\ }\textbf
  {\bibinfo {volume} {189}},\ \bibinfo {pages} {269} (\bibinfo {year}
  {2011})}\BibitemShut {NoStop}%
\bibitem [{\citenamefont {Cardoso}\ \emph {et~al.}(2012)\citenamefont
  {Cardoso}, \citenamefont {Gualtieri}, \citenamefont {Herdeiro}, \citenamefont
  {Sperhake}, \citenamefont {Chesler} \emph {et~al.}}]{Cardoso:2012qm}%
  \BibitemOpen
  \bibfield  {author} {\bibinfo {author} {\bibfnamefont {V.}~\bibnamefont
  {Cardoso}}, \bibinfo {author} {\bibfnamefont {L.}~\bibnamefont {Gualtieri}},
  \bibinfo {author} {\bibfnamefont {C.}~\bibnamefont {Herdeiro}}, \bibinfo
  {author} {\bibfnamefont {U.}~\bibnamefont {Sperhake}}, \bibinfo {author}
  {\bibfnamefont {P.~M.}\ \bibnamefont {Chesler}},  \emph {et~al.},\ }\href
  {\doibase 10.1088/0264-9381/29/24/244001} {\bibfield  {journal} {\bibinfo
  {journal} {Class.Quant.Grav.}\ }\textbf {\bibinfo {volume} {29}},\ \bibinfo
  {pages} {244001} (\bibinfo {year} {2012})},\ \Eprint
  {http://arxiv.org/abs/1201.5118} {arXiv:1201.5118 [hep-th]} \BibitemShut
  {NoStop}%
\bibitem [{\citenamefont {Teukolsky}(1972)}]{Teukolsky:1972my}%
  \BibitemOpen
  \bibfield  {author} {\bibinfo {author} {\bibfnamefont {S.~A.}\ \bibnamefont
  {Teukolsky}},\ }\href {\doibase 10.1103/PhysRevLett.29.1114} {\bibfield
  {journal} {\bibinfo  {journal} {Phys. Rev. Lett.}\ }\textbf {\bibinfo
  {volume} {29}},\ \bibinfo {pages} {1114} (\bibinfo {year}
  {1972})}\BibitemShut {NoStop}%
\bibitem [{\citenamefont {Teukolsky}(1973)}]{Teukolsky:1973ap}%
  \BibitemOpen
  \bibfield  {author} {\bibinfo {author} {\bibfnamefont {S.~A.}\ \bibnamefont
  {Teukolsky}},\ }\href {\doibase 10.1086/152444} {\bibfield  {journal}
  {\bibinfo  {journal} {Astrophys. J.}\ }\textbf {\bibinfo {volume} {185}},\
  \bibinfo {pages} {635} (\bibinfo {year} {1973})}\BibitemShut {NoStop}%
\bibitem [{\citenamefont {Berti}\ \emph
  {et~al.}(2006{\natexlab{a}})\citenamefont {Berti}, \citenamefont {Cardoso},\
  and\ \citenamefont {Casals}}]{Berti:2005gp}%
  \BibitemOpen
  \bibfield  {author} {\bibinfo {author} {\bibfnamefont {E.}~\bibnamefont
  {Berti}}, \bibinfo {author} {\bibfnamefont {V.}~\bibnamefont {Cardoso}}, \
  and\ \bibinfo {author} {\bibfnamefont {M.}~\bibnamefont {Casals}},\ }\href
  {\doibase 10.1103/PhysRevD.73.024013} {\bibfield  {journal} {\bibinfo
  {journal} {Phys. Rev.}\ }\textbf {\bibinfo {volume} {D73}},\ \bibinfo {pages}
  {024013} (\bibinfo {year} {2006}{\natexlab{a}})},\ \Eprint
  {http://arxiv.org/abs/gr-qc/0511111} {arXiv:gr-qc/0511111} \BibitemShut
  {NoStop}%
\bibitem [{\citenamefont {Nollert}\ and\ \citenamefont
  {Price}(1999)}]{Nollert:1998ys}%
  \BibitemOpen
  \bibfield  {author} {\bibinfo {author} {\bibfnamefont {H.-P.}\ \bibnamefont
  {Nollert}}\ and\ \bibinfo {author} {\bibfnamefont {R.~H.}\ \bibnamefont
  {Price}},\ }\href {\doibase 10.1063/1.532698} {\bibfield  {journal} {\bibinfo
   {journal} {J. Math. Phys.}\ }\textbf {\bibinfo {volume} {40}},\ \bibinfo
  {pages} {980} (\bibinfo {year} {1999})},\ \Eprint
  {http://arxiv.org/abs/gr-qc/9810074} {arXiv:gr-qc/9810074} \BibitemShut
  {NoStop}%
\bibitem [{\citenamefont {Berti}\ \emph
  {et~al.}(2006{\natexlab{b}})\citenamefont {Berti}, \citenamefont {Cardoso},\
  and\ \citenamefont {Will}}]{Berti:2006hb}%
  \BibitemOpen
  \bibfield  {author} {\bibinfo {author} {\bibfnamefont {E.}~\bibnamefont
  {Berti}}, \bibinfo {author} {\bibfnamefont {V.}~\bibnamefont {Cardoso}}, \
  and\ \bibinfo {author} {\bibfnamefont {C.~M.}\ \bibnamefont {Will}},\ }\href
  {\doibase 10.1063/1.2348047} {\bibfield  {journal} {\bibinfo  {journal} {AIP
  Conf. Proc.}\ }\textbf {\bibinfo {volume} {848}},\ \bibinfo {pages} {687}
  (\bibinfo {year} {2006}{\natexlab{b}})},\ \Eprint
  {http://arxiv.org/abs/gr-qc/0601077} {arXiv:gr-qc/0601077} \BibitemShut
  {NoStop}%
\bibitem [{\citenamefont {Dorband}\ \emph {et~al.}(2006)\citenamefont
  {Dorband}, \citenamefont {Berti}, \citenamefont {Diener}, \citenamefont
  {Schnetter},\ and\ \citenamefont {Tiglio}}]{Dorband:2006gg}%
  \BibitemOpen
  \bibfield  {author} {\bibinfo {author} {\bibfnamefont {E.~N.}\ \bibnamefont
  {Dorband}}, \bibinfo {author} {\bibfnamefont {E.}~\bibnamefont {Berti}},
  \bibinfo {author} {\bibfnamefont {P.}~\bibnamefont {Diener}}, \bibinfo
  {author} {\bibfnamefont {E.}~\bibnamefont {Schnetter}}, \ and\ \bibinfo
  {author} {\bibfnamefont {M.}~\bibnamefont {Tiglio}},\ }\href {\doibase
  10.1103/PhysRevD.74.084028} {\bibfield  {journal} {\bibinfo  {journal} {Phys.
  Rev.}\ }\textbf {\bibinfo {volume} {D74}},\ \bibinfo {pages} {084028}
  (\bibinfo {year} {2006})},\ \Eprint {http://arxiv.org/abs/gr-qc/0608091}
  {arXiv:gr-qc/0608091} \BibitemShut {NoStop}%
\bibitem [{\citenamefont {Sun}\ and\ \citenamefont {Price}(1988)}]{Sun:1988tz}%
  \BibitemOpen
  \bibfield  {author} {\bibinfo {author} {\bibfnamefont {Y.}~\bibnamefont
  {Sun}}\ and\ \bibinfo {author} {\bibfnamefont {R.~H.}\ \bibnamefont
  {Price}},\ }\href {\doibase 10.1103/PhysRevD.38.1040} {\bibfield  {journal}
  {\bibinfo  {journal} {Phys. Rev.}\ }\textbf {\bibinfo {volume} {D38}},\
  \bibinfo {pages} {1040} (\bibinfo {year} {1988})}\BibitemShut {NoStop}%
\bibitem [{\citenamefont {Hadar}\ and\ \citenamefont
  {Kol}(2011)}]{Hadar:2009ip}%
  \BibitemOpen
  \bibfield  {author} {\bibinfo {author} {\bibfnamefont {S.}~\bibnamefont
  {Hadar}}\ and\ \bibinfo {author} {\bibfnamefont {B.}~\bibnamefont {Kol}},\
  }\href {\doibase 10.1103/PhysRevD.84.044019} {\bibfield  {journal} {\bibinfo
  {journal} {Phys. Rev.}\ }\textbf {\bibinfo {volume} {D84}},\ \bibinfo {pages}
  {044019} (\bibinfo {year} {2011})},\ \Eprint {http://arxiv.org/abs/0911.3899}
  {arXiv:0911.3899 [gr-qc]} \BibitemShut {NoStop}%
\bibitem [{\citenamefont {Hadar}\ \emph {et~al.}(2011)\citenamefont {Hadar},
  \citenamefont {Kol}, \citenamefont {Berti},\ and\ \citenamefont
  {Cardoso}}]{Hadar:2011vj}%
  \BibitemOpen
  \bibfield  {author} {\bibinfo {author} {\bibfnamefont {S.}~\bibnamefont
  {Hadar}}, \bibinfo {author} {\bibfnamefont {B.}~\bibnamefont {Kol}}, \bibinfo
  {author} {\bibfnamefont {E.}~\bibnamefont {Berti}}, \ and\ \bibinfo {author}
  {\bibfnamefont {V.}~\bibnamefont {Cardoso}},\ }\href {\doibase
  10.1103/PhysRevD.84.047501} {\bibfield  {journal} {\bibinfo  {journal} {Phys.
  Rev.}\ }\textbf {\bibinfo {volume} {D84}},\ \bibinfo {pages} {047501}
  (\bibinfo {year} {2011})},\ \Eprint {http://arxiv.org/abs/1105.3861}
  {arXiv:1105.3861 [gr-qc]} \BibitemShut {NoStop}%
\bibitem [{\citenamefont {Sasaki}\ and\ \citenamefont
  {Nakamura}(1982)}]{Sasaki:1981sx}%
  \BibitemOpen
  \bibfield  {author} {\bibinfo {author} {\bibfnamefont {M.}~\bibnamefont
  {Sasaki}}\ and\ \bibinfo {author} {\bibfnamefont {T.}~\bibnamefont
  {Nakamura}},\ }\href {\doibase 10.1143/PTP.67.1788} {\bibfield  {journal}
  {\bibinfo  {journal} {Prog. Theor. Phys.}\ }\textbf {\bibinfo {volume}
  {67}},\ \bibinfo {pages} {1788} (\bibinfo {year} {1982})}\BibitemShut
  {NoStop}%
\bibitem [{\citenamefont {Regge}\ and\ \citenamefont
  {Wheeler}(1957)}]{Regge:1957rw}%
  \BibitemOpen
  \bibfield  {author} {\bibinfo {author} {\bibfnamefont {T.}~\bibnamefont
  {Regge}}\ and\ \bibinfo {author} {\bibfnamefont {J.~A.}\ \bibnamefont
  {Wheeler}},\ }\href {\doibase 10.1103/PhysRev.108.1063} {\bibfield  {journal}
  {\bibinfo  {journal} {Phys. Rev.}\ }\textbf {\bibinfo {volume} {108}},\
  \bibinfo {pages} {1063} (\bibinfo {year} {1957})}\BibitemShut {NoStop}%
\bibitem [{\citenamefont {Zerilli}(1970)}]{Zerilli:1971wd}%
  \BibitemOpen
  \bibfield  {author} {\bibinfo {author} {\bibfnamefont {F.~J.}\ \bibnamefont
  {Zerilli}},\ }\href {\doibase 10.1103/PhysRevD.2.2141} {\bibfield  {journal}
  {\bibinfo  {journal} {Phys. Rev.}\ }\textbf {\bibinfo {volume} {D2}},\
  \bibinfo {pages} {2141} (\bibinfo {year} {1970})}\BibitemShut {NoStop}%
\bibitem [{\citenamefont {Mano}\ \emph
  {et~al.}(1996{\natexlab{a}})\citenamefont {Mano}, \citenamefont {Suzuki},\
  and\ \citenamefont {Takasugi}}]{Mano:1996vt}%
  \BibitemOpen
  \bibfield  {author} {\bibinfo {author} {\bibfnamefont {S.}~\bibnamefont
  {Mano}}, \bibinfo {author} {\bibfnamefont {H.}~\bibnamefont {Suzuki}}, \ and\
  \bibinfo {author} {\bibfnamefont {E.}~\bibnamefont {Takasugi}},\ }\href
  {\doibase 10.1143/PTP.95.1079} {\bibfield  {journal} {\bibinfo  {journal}
  {Prog. Theor. Phys.}\ }\textbf {\bibinfo {volume} {95}},\ \bibinfo {pages}
  {1079} (\bibinfo {year} {1996}{\natexlab{a}})},\ \Eprint
  {http://arxiv.org/abs/gr-qc/9603020} {arXiv:gr-qc/9603020} \BibitemShut
  {NoStop}%
\bibitem [{\citenamefont {Mano}\ \emph
  {et~al.}(1996{\natexlab{b}})\citenamefont {Mano}, \citenamefont {Suzuki},\
  and\ \citenamefont {Takasugi}}]{Mano:1996mf}%
  \BibitemOpen
  \bibfield  {author} {\bibinfo {author} {\bibfnamefont {S.}~\bibnamefont
  {Mano}}, \bibinfo {author} {\bibfnamefont {H.}~\bibnamefont {Suzuki}}, \ and\
  \bibinfo {author} {\bibfnamefont {E.}~\bibnamefont {Takasugi}},\ }\href
  {\doibase 10.1143/PTP.96.549} {\bibfield  {journal} {\bibinfo  {journal}
  {Prog. Theor. Phys.}\ }\textbf {\bibinfo {volume} {96}},\ \bibinfo {pages}
  {549} (\bibinfo {year} {1996}{\natexlab{b}})},\ \Eprint
  {http://arxiv.org/abs/gr-qc/9605057} {arXiv:gr-qc/9605057} \BibitemShut
  {NoStop}%
\bibitem [{rdw()}]{rdweb}%
  \BibitemOpen
  \href@noop {} {}\bibinfo {note} {{Webpage with Mathematica notebooks and
  numerical quasinormal mode Tables: \\
  \url{http://www.phy.olemiss.edu/~berti/qnms.html} \\
  \url{http://gamow.ist.utl.pt/~vitor/ringdown.html} }}\BibitemShut {NoStop}%
\bibitem [{\citenamefont {Mano}\ and\ \citenamefont
  {Takasugi}(1997)}]{Mano:1996gn}%
  \BibitemOpen
  \bibfield  {author} {\bibinfo {author} {\bibfnamefont {S.}~\bibnamefont
  {Mano}}\ and\ \bibinfo {author} {\bibfnamefont {E.}~\bibnamefont
  {Takasugi}},\ }\href {\doibase 10.1143/PTP.97.213} {\bibfield  {journal}
  {\bibinfo  {journal} {Prog. Theor. Phys.}\ }\textbf {\bibinfo {volume}
  {97}},\ \bibinfo {pages} {213} (\bibinfo {year} {1997})},\ \Eprint
  {http://arxiv.org/abs/gr-qc/9611014} {arXiv:gr-qc/9611014} \BibitemShut
  {NoStop}%
\bibitem [{\citenamefont {Leaver}(1985)}]{Leaver:1985ax}%
  \BibitemOpen
  \bibfield  {author} {\bibinfo {author} {\bibfnamefont {E.~W.}\ \bibnamefont
  {Leaver}},\ }\href {\doibase 10.1098/rspa.1985.0119} {\bibfield  {journal}
  {\bibinfo  {journal} {Proc. R. Soc. London, Ser. A}\ }\textbf {\bibinfo
  {volume} {402}},\ \bibinfo {pages} {285} (\bibinfo {year}
  {1985})}\BibitemShut {NoStop}%
\bibitem [{\citenamefont {Sasaki}\ and\ \citenamefont
  {Tagoshi}(2003)}]{Sasaki:2003xr}%
  \BibitemOpen
  \bibfield  {author} {\bibinfo {author} {\bibfnamefont {M.}~\bibnamefont
  {Sasaki}}\ and\ \bibinfo {author} {\bibfnamefont {H.}~\bibnamefont
  {Tagoshi}},\ }\href {http://www.livingreviews.org/lrr-2003-6} {\bibfield
  {journal} {\bibinfo  {journal} {Living Rev. Relativity}\ }\textbf {\bibinfo
  {volume} {6}} (\bibinfo {year} {2003})},\ \Eprint
  {http://arxiv.org/abs/gr-qc/0306120} {arXiv:gr-qc/0306120} \BibitemShut
  {NoStop}%
\bibitem [{\citenamefont {Chandrasekhar}(1983)}]{Chandrasekhar:1985kt}%
  \BibitemOpen
  \bibfield  {author} {\bibinfo {author} {\bibfnamefont {S.}~\bibnamefont
  {Chandrasekhar}},\ }\href@noop {} {\emph {\bibinfo {title} {The Mathematical
  Theory of Black Holes}}}\ (\bibinfo  {publisher} {Clarendon Press},\ \bibinfo
  {address} {Oxford, U.K.},\ \bibinfo {year} {1983})\BibitemShut {NoStop}%
\bibitem [{\citenamefont {Davis}\ \emph {et~al.}(1971)\citenamefont {Davis},
  \citenamefont {Ruffini}, \citenamefont {Press},\ and\ \citenamefont
  {Price}}]{Davis:1971gg}%
  \BibitemOpen
  \bibfield  {author} {\bibinfo {author} {\bibfnamefont {M.}~\bibnamefont
  {Davis}}, \bibinfo {author} {\bibfnamefont {R.~J.}\ \bibnamefont {Ruffini}},
  \bibinfo {author} {\bibfnamefont {W.~H.}\ \bibnamefont {Press}}, \ and\
  \bibinfo {author} {\bibfnamefont {R.~H.}\ \bibnamefont {Price}},\ }\href
  {\doibase 10.1103/PhysRevLett.27.1466} {\bibfield  {journal} {\bibinfo
  {journal} {Phys. Rev. Lett.}\ }\textbf {\bibinfo {volume} {27}},\ \bibinfo
  {pages} {1466} (\bibinfo {year} {1971})}\BibitemShut {NoStop}%
\bibitem [{\citenamefont {Ferrari}\ and\ \citenamefont
  {Ruffini}(1981)}]{Ferrari:1981dh}%
  \BibitemOpen
  \bibfield  {author} {\bibinfo {author} {\bibfnamefont {V.}~\bibnamefont
  {Ferrari}}\ and\ \bibinfo {author} {\bibfnamefont {R.}~\bibnamefont
  {Ruffini}},\ }\href {\doibase 10.1016/0370-2693(81)90930-8} {\bibfield
  {journal} {\bibinfo  {journal} {Phys.Lett.}\ }\textbf {\bibinfo {volume}
  {B98}},\ \bibinfo {pages} {381} (\bibinfo {year} {1981})}\BibitemShut
  {NoStop}%
\bibitem [{\citenamefont {Lousto}\ and\ \citenamefont
  {Price}(1997)}]{Lousto:1996sx}%
  \BibitemOpen
  \bibfield  {author} {\bibinfo {author} {\bibfnamefont {C.~O.}\ \bibnamefont
  {Lousto}}\ and\ \bibinfo {author} {\bibfnamefont {R.~H.}\ \bibnamefont
  {Price}},\ }\href {\doibase 10.1103/PhysRevD.55.2124} {\bibfield  {journal}
  {\bibinfo  {journal} {Phys. Rev.}\ }\textbf {\bibinfo {volume} {D55}},\
  \bibinfo {pages} {2124} (\bibinfo {year} {1997})},\ \Eprint
  {http://arxiv.org/abs/gr-qc/9609012} {arXiv:gr-qc/9609012} \BibitemShut
  {NoStop}%
\bibitem [{\citenamefont {Cardoso}\ and\ \citenamefont
  {Lemos}(2002)}]{Cardoso:2002ay}%
  \BibitemOpen
  \bibfield  {author} {\bibinfo {author} {\bibfnamefont {V.}~\bibnamefont
  {Cardoso}}\ and\ \bibinfo {author} {\bibfnamefont {J.~P.~S.}\ \bibnamefont
  {Lemos}},\ }\href {\doibase 10.1016/S0370-2693(02)01961-5} {\bibfield
  {journal} {\bibinfo  {journal} {Phys. Lett.}\ }\textbf {\bibinfo {volume}
  {B538}},\ \bibinfo {pages} {1} (\bibinfo {year} {2002})},\ \Eprint
  {http://arxiv.org/abs/gr-qc/0202019} {arXiv:gr-qc/0202019} \BibitemShut
  {NoStop}%
\bibitem [{\citenamefont {Cardoso}\ and\ \citenamefont
  {Lemos}(2003{\natexlab{a}})}]{Cardoso:2002yj}%
  \BibitemOpen
  \bibfield  {author} {\bibinfo {author} {\bibfnamefont {V.}~\bibnamefont
  {Cardoso}}\ and\ \bibinfo {author} {\bibfnamefont {J.~P.~S.}\ \bibnamefont
  {Lemos}},\ }\href {\doibase 10.1023/A:1022301412348} {\bibfield  {journal}
  {\bibinfo  {journal} {Gen. Rel. Grav.}\ }\textbf {\bibinfo {volume} {35}},\
  \bibinfo {pages} {327} (\bibinfo {year} {2003}{\natexlab{a}})},\ \Eprint
  {http://arxiv.org/abs/gr-qc/0207009} {arXiv:gr-qc/0207009} \BibitemShut
  {NoStop}%
\bibitem [{\citenamefont {Kodama}\ and\ \citenamefont
  {Ishibashi}(2003)}]{Kodama:2003jz}%
  \BibitemOpen
  \bibfield  {author} {\bibinfo {author} {\bibfnamefont {H.}~\bibnamefont
  {Kodama}}\ and\ \bibinfo {author} {\bibfnamefont {A.}~\bibnamefont
  {Ishibashi}},\ }\href {\doibase 10.1143/PTP.110.701} {\bibfield  {journal}
  {\bibinfo  {journal} {Prog. Theor. Phys.}\ }\textbf {\bibinfo {volume}
  {110}},\ \bibinfo {pages} {701} (\bibinfo {year} {2003})},\ \Eprint
  {http://arxiv.org/abs/hep-th/0305147} {arXiv:hep-th/0305147} \BibitemShut
  {NoStop}%
\bibitem [{\citenamefont {Berti}\ \emph {et~al.}(2010)\citenamefont {Berti},
  \citenamefont {Cardoso}, \citenamefont {Hinderer}, \citenamefont {Lemos},
  \citenamefont {Pretorius}, \citenamefont {Sperhake},\ and\ \citenamefont
  {Yunes}}]{Berti:2010ce}%
  \BibitemOpen
  \bibfield  {author} {\bibinfo {author} {\bibfnamefont {E.}~\bibnamefont
  {Berti}}, \bibinfo {author} {\bibfnamefont {V.}~\bibnamefont {Cardoso}},
  \bibinfo {author} {\bibfnamefont {T.}~\bibnamefont {Hinderer}}, \bibinfo
  {author} {\bibfnamefont {M.}~\bibnamefont {Lemos}}, \bibinfo {author}
  {\bibfnamefont {F.}~\bibnamefont {Pretorius}}, \bibinfo {author}
  {\bibfnamefont {U.}~\bibnamefont {Sperhake}}, \ and\ \bibinfo {author}
  {\bibfnamefont {N.}~\bibnamefont {Yunes}},\ }\href {\doibase
  10.1103/PhysRevD.81.104048} {\bibfield  {journal} {\bibinfo  {journal} {Phys.
  Rev.}\ }\textbf {\bibinfo {volume} {D81}},\ \bibinfo {pages} {104048}
  (\bibinfo {year} {2010})},\ \Eprint {http://arxiv.org/abs/1003.0812}
  {arXiv:1003.0812 [gr-qc]} \BibitemShut {NoStop}%
\bibitem [{\citenamefont {Mitsou}(2011)}]{Mitsou:2010jv}%
  \BibitemOpen
  \bibfield  {author} {\bibinfo {author} {\bibfnamefont {E.}~\bibnamefont
  {Mitsou}},\ }\href {\doibase 10.1103/PhysRevD.83.044039} {\bibfield
  {journal} {\bibinfo  {journal} {Phys.Rev.}\ }\textbf {\bibinfo {volume}
  {D83}},\ \bibinfo {pages} {044039} (\bibinfo {year} {2011})},\ \Eprint
  {http://arxiv.org/abs/1012.2028} {arXiv:1012.2028 [gr-qc]} \BibitemShut
  {NoStop}%
\bibitem [{\citenamefont {Cardoso}\ and\ \citenamefont
  {Lemos}(2003{\natexlab{b}})}]{Cardoso:2002jr}%
  \BibitemOpen
  \bibfield  {author} {\bibinfo {author} {\bibfnamefont {V.}~\bibnamefont
  {Cardoso}}\ and\ \bibinfo {author} {\bibfnamefont {J.~P.~S.}\ \bibnamefont
  {Lemos}},\ }\href {\doibase 10.1103/PhysRevD.67.084005} {\bibfield  {journal}
  {\bibinfo  {journal} {Phys. Rev.}\ }\textbf {\bibinfo {volume} {D67}},\
  \bibinfo {pages} {084005} (\bibinfo {year} {2003}{\natexlab{b}})},\ \Eprint
  {http://arxiv.org/abs/gr-qc/0211094} {arXiv:gr-qc/0211094} \BibitemShut
  {NoStop}%
\bibitem [{\citenamefont {Detweiler}\ and\ \citenamefont
  {Szedenits}(1979)}]{Detweiler:1979xr}%
  \BibitemOpen
  \bibfield  {author} {\bibinfo {author} {\bibfnamefont {S.~L.}\ \bibnamefont
  {Detweiler}}\ and\ \bibinfo {author} {\bibfnamefont {J.}~\bibnamefont
  {Szedenits}, \bibfnamefont {Eugene}},\ }\href {\doibase 10.1086/157182}
  {\bibfield  {journal} {\bibinfo  {journal} {Astrophys. J.}\ }\textbf
  {\bibinfo {volume} {231}},\ \bibinfo {pages} {211} (\bibinfo {year}
  {1979})}\BibitemShut {NoStop}%
\bibitem [{\citenamefont {Andersson}\ and\ \citenamefont
  {Glampedakis}(2000)}]{Andersson:1999wj}%
  \BibitemOpen
  \bibfield  {author} {\bibinfo {author} {\bibfnamefont {N.}~\bibnamefont
  {Andersson}}\ and\ \bibinfo {author} {\bibfnamefont {K.}~\bibnamefont
  {Glampedakis}},\ }\href {\doibase 10.1103/PhysRevLett.84.4537} {\bibfield
  {journal} {\bibinfo  {journal} {Phys.Rev.Lett.}\ }\textbf {\bibinfo {volume}
  {84}},\ \bibinfo {pages} {4537} (\bibinfo {year} {2000})},\ \Eprint
  {http://arxiv.org/abs/gr-qc/9909050} {arXiv:gr-qc/9909050 [gr-qc]}
  \BibitemShut {NoStop}%
\bibitem [{\citenamefont {Cardoso}(2004)}]{Cardoso:2004hh}%
  \BibitemOpen
  \bibfield  {author} {\bibinfo {author} {\bibfnamefont {V.}~\bibnamefont
  {Cardoso}},\ }\href {\doibase 10.1103/PhysRevD.70.127502} {\bibfield
  {journal} {\bibinfo  {journal} {Phys. Rev.}\ }\textbf {\bibinfo {volume}
  {D70}},\ \bibinfo {pages} {127502} (\bibinfo {year} {2004})},\ \Eprint
  {http://arxiv.org/abs/gr-qc/0411048} {arXiv:gr-qc/0411048} \BibitemShut
  {NoStop}%
\bibitem [{\citenamefont {Berti}\ \emph
  {et~al.}(2006{\natexlab{c}})\citenamefont {Berti}, \citenamefont {Cardoso},\
  and\ \citenamefont {Will}}]{Berti:2005ys}%
  \BibitemOpen
  \bibfield  {author} {\bibinfo {author} {\bibfnamefont {E.}~\bibnamefont
  {Berti}}, \bibinfo {author} {\bibfnamefont {V.}~\bibnamefont {Cardoso}}, \
  and\ \bibinfo {author} {\bibfnamefont {C.~M.}\ \bibnamefont {Will}},\ }\href
  {\doibase 10.1103/PhysRevD.73.064030} {\bibfield  {journal} {\bibinfo
  {journal} {Phys. Rev.}\ }\textbf {\bibinfo {volume} {D73}},\ \bibinfo {pages}
  {064030} (\bibinfo {year} {2006}{\natexlab{c}})},\ \Eprint
  {http://arxiv.org/abs/gr-qc/0512160} {arXiv:gr-qc/0512160} \BibitemShut
  {NoStop}%
\bibitem [{\citenamefont {Yang}\ \emph {et~al.}(2013)\citenamefont {Yang},
  \citenamefont {Zhang}, \citenamefont {Zimmerman}, \citenamefont {Nichols},
  \citenamefont {Berti} \emph {et~al.}}]{Yang:2012pj}%
  \BibitemOpen
  \bibfield  {author} {\bibinfo {author} {\bibfnamefont {H.}~\bibnamefont
  {Yang}}, \bibinfo {author} {\bibfnamefont {F.}~\bibnamefont {Zhang}},
  \bibinfo {author} {\bibfnamefont {A.}~\bibnamefont {Zimmerman}}, \bibinfo
  {author} {\bibfnamefont {D.~A.}\ \bibnamefont {Nichols}}, \bibinfo {author}
  {\bibfnamefont {E.}~\bibnamefont {Berti}},  \emph {et~al.},\ }\href {\doibase
  10.1103/PhysRevD.87.041502} {\bibfield  {journal} {\bibinfo  {journal}
  {Phys.Rev.}\ }\textbf {\bibinfo {volume} {D87}},\ \bibinfo {pages} {041502}
  (\bibinfo {year} {2013})},\ \Eprint {http://arxiv.org/abs/1212.3271}
  {arXiv:1212.3271 [gr-qc]} \BibitemShut {NoStop}%
\end{thebibliography}
%

\end{document}